\documentclass{ws-p8-50x6-00}

\newcommand{\ba}[1]{\begin{eqnarray} \label{(#1)}}
\newcommand{\ea}{\end{eqnarray}}
\newcommand{\nn}{\nonumber}

\newcommand{\AmS}{{\protect\the\textfont2
  A\kern-.1667em\lower.5ex\hbox{M}\kern-.125emS}}

\def \znbb {$0\nu\beta\beta$}

\def\be{\begin{equation}}
\def\ee{\end{equation}}
\def\bea{\begin{eqnarray}}
\def\eea{\end{eqnarray}}

\def \gsim {~\mbox{${}^> \hspace*{-9pt} _\sim$}~}

\begin{document}

\title{Search for Neutrino Mass and Dark Matter in Underground Experiments}

\author{H.V. Klapdor-Kleingrothaus}

\address{Max-Planck-Institut f\"ur Kernphysik,\\ 
P.O. Box 10 39 80, D-69029 Heidelberg, Germany\\ 
Spokesman of HEIDELBERG-MOSCOW and GENIUS Collaborations\\
E-mail: klapdor@gustav.mpi-hd.mpg,\\
 Home-page: http://www.mpi-hd.mpg.de.non\_acc/}

\maketitle


\abstracts{
	Search for the neutrino mass and for cold dark matter 
	in the Universe are at present two of the most exiting fields  
	of particle physics and cosmology. 
	This lecture will restrict itself on the search 
	for neutralinos as cold dark matter, and for the absolute 
	scale of 
	the masses of neutrinos, 
	which are the favoured hot dark matter candidates.}


\section{Introduction}

	In this lecture we shall discuss two central problems 
	of particle physics and cosmology. 
	The neutrino mass is one of the key quantities 
	in grand unified theories, and at the same 
	time candidate for hot dark matter in the Universe. 
	Supersymmetry is regarded as the most natural extension 
	of present particle physics theories. 
	The lighest SUSY particle, usually assumed to be the neutralino, 
	is the favored candidate for cold dark matter. 
	We shall discuss here the search for the neutrino mass, 
	concentrating on double beta decay, 
	and the search for neutralinos, in underground experiments.

	The neutrino oscillation
	interpretation of the atmospheric and solar neutrino data,  
	deliver a strong indication for a non-vanishing neutrino mass.
	While such kind of experiments yields information on the difference of 
	squared neutrino mass eigenvalues and on mixing angles, 
	they cannot fix the 
	absolute scale of the neutrino mass.
	Information from double beta decay experiments is indispensable to
	solve this question 
\cite{KK-Sark01-Ev,KKPS,KKPS-01,KK60Y}.
	Another important problem is that of the fundamental
	character of the neutrino, whether it is a Dirac or a Majorana
	particle 
\cite{Majorana37,Rac37}.
	Neutrinoless double beta decay could answer also this question. 
	Perhaps the main question, which can be investigated 
	by double beta decay with high sensitivity, is  
	that of lepton number conservation or non-conservation.

	The physics motivations to search for dark matter 
	are manyfold.

	Recent investigation of the cosmic microwave background 
	radiation (MAXIMA, BOOMERANG, DASI) together with 
	large scale structure results fix 
	$\Omega_\Lambda~+~\Omega_m~\sim~~1$, where 
	$\Omega_\Lambda~$ ($=\rho_{\Lambda}/\rho_c$) stands the 
	for dark energy and $\Omega_m$ for matter. 
	With the early nucleosynthesis constraint of 
	$\Omega_{bar}\sim$0.04 the need 
	for non-baryonic dark matter is evident. 
	This is true even for our galaxy, since MACHOs represent 
	only a small fraction of galactic dark matter%
\cite{Freez}.

	Although there exist other candidates such as axions%
\cite{Axions}, 
	gravitinos, etc., 
	neutralinos seem to be the favored candidates at present. 

	Hot dark matter, according to CMB and LSS (Redshift-Survey) 
	results still contributes up to 38$\%$ of the dark matter%
\cite{Teg00}.
	This corresponds to a sum of neutrino masses 
	$<$5.5\,eV. 

\begin{figure}[h]

\vspace{-0.4cm}
\begin{center}
\epsfig{file=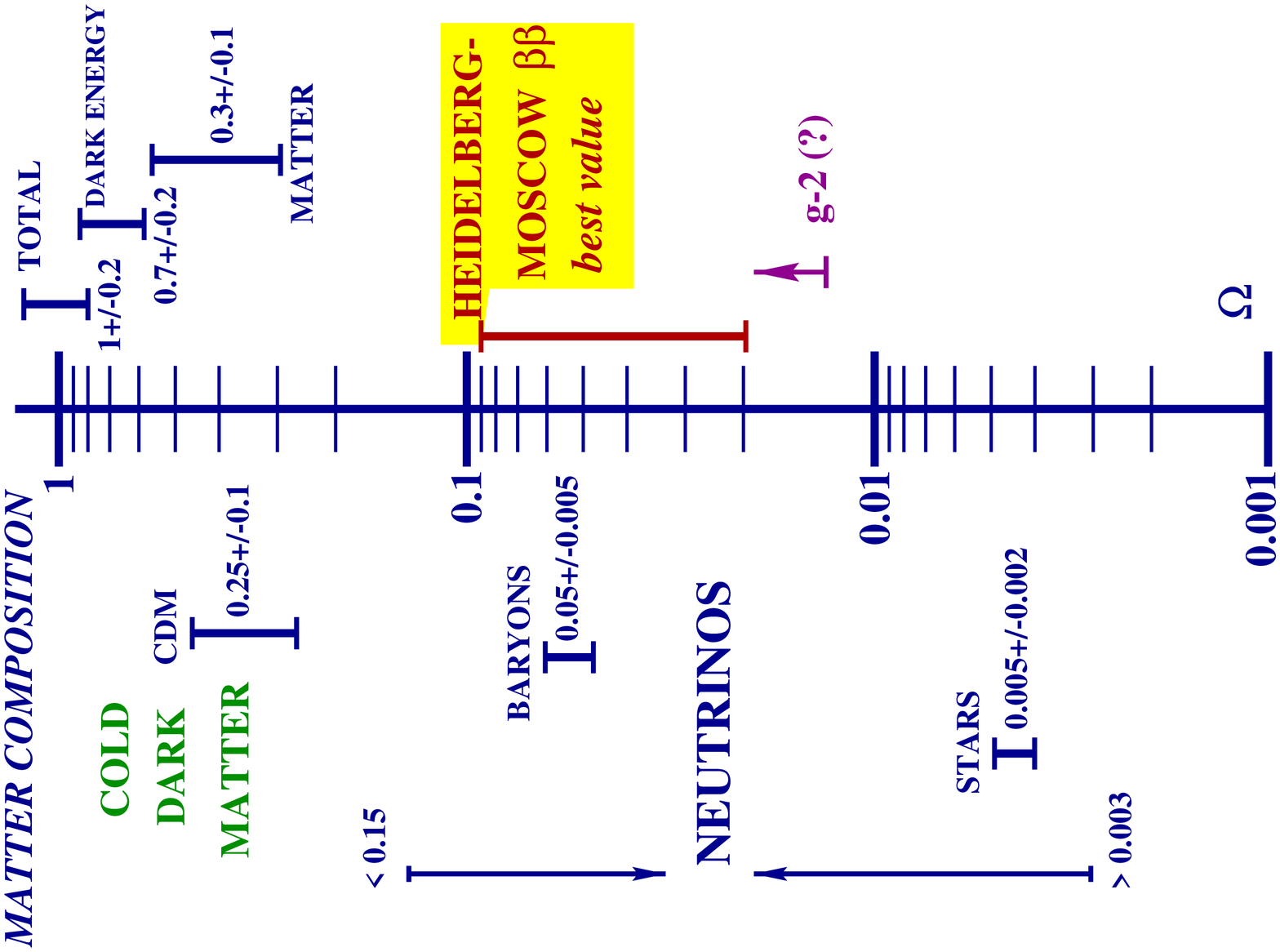,width=75mm, height=70mm, angle=-90}
\end{center}
%
\caption[]{
	Our present understanding of the mass  
	composition of the Universe.
\label{fig:DarkMatter}}
\end{figure}

	Our present picture of the mass/energy distribution 
	in the Universe is as given in Fig.%
\ref{fig:DarkMatter}.

\vspace{-0.9cm}
	In section 2, we discuss double beta decay, 
	including in section 2.2 the expectation from neutrino oscillation 
	experiments, and in section 2.3 the recent evidence 
	for the neutrinoless decay mode from the HEIDELBERG-MOSCOW 
	experiment. In section 2.4 we discuss the future of the field.

	In section 3 we shall discuss dark matter search. 
	We start in 3.1 with the expectation from SUSY models 
	and then discuss in section 3.2 the present status of 
	cold dark matter experiments - including the important indication 
	from the DAMA experiment - and its future.
	In section 3.3 we give some comments to neutrinos 
	as hot dark matter.
	In section 4 we give a conclusion.



\section{Neutrino Masses and Double Beta Decay}

\subsection{General}

	Double beta decay, the rarest known nuclear decay process, can occur
	in different modes:

\begin{tabbing}
\hspace*{4.5cm} \= \hspace*{7.9cm} \= \kill
\qquad $ 2\nu\beta\beta- \rm decay:$ \> $ A(Z,N) \rightarrow
A(Z\!+\!2,N\!-\!2) + 2e^- + 2\bar{\nu}_e$ \hspace*{1.05cm}(1)\\[-0ex]
\qquad $ 0\nu\beta\beta- \rm decay:$ \> $ A(Z,N) \rightarrow
A(Z\!+\!2,N\!-\!2) + 2e^-$ \hspace*{2cm}(2) \\[-0ex]
\qquad $ 0\nu(2)\chi\beta\beta- \rm decay:$ \> $ A(Z,N) \rightarrow
A(Z\!+\!2,N\!-\!2) + 2e^- + (2)\chi$ \hspace*{0.9cm}(3) \\[-0ex]
\end{tabbing}

\vspace{-0.3cm}
	While the two-neutrino mode (1)
	is allowed by the Standard Model of particle physics, 
	the neutrinoless mode 
	(\znbb) (2) requires violation of lepton number 
	($\Delta$L=2). 
	This mode is possible only, if the neutrino 
	is a Majorana particle, i.e. the neutrino is its own antiparticle 
	(E. Majorana
\cite{Majorana37},  
	G. Racah 
\cite{Rac37},    
	 for subsequent works we refer to 
\cite{Mcl57,Case57,Ahl96}, 
	 for some reviews see   	
\cite{Doi85,Mut88,Gro89/90,Kla95/98,KK60Y,Vog2001}). 
	First calculations of \znbb~ decay based 
	on the Majorana theory have been done by W.H. Furry 
\cite{Fur39}.

	Neutrinoless double beta decay can not only probe a
	Majorana neutrino mass, but various new physics scenarios beyond the
	Standard Model, such as R-parity violating supersymmetric models, 
	R-parity conserving SUSY models,  
	leptoquarks, 
	violation of Lorentz-invariance,  
	and compositeness 
	(for a review see 
	\cite{KK60Y,KK-LeptBar98,KK-SprTracts00}). 
	Any theory containing lepton number violating
	interactions can in principle lead to this process allowing to obtain
	information on the specific underlying theory.
	The experimental signature of the neutrinoless mode is a peak at the
	Q-value of the decay.

	The unique feature of neutrinoless double beta decay is that 
	a measured half-life allows to deduce information on 
	the effective Majorana neutrino mass
$\langle m \rangle $ , 
	which is a superposition of neutrino mass eigenstates:
\cite{Doi85,Mut88}
	\bea 
[T^{0\nu}_{1/2}(0^+_i \rightarrow 0^+_f)]^{-1}= C_{mm} 
\frac{\langle m \rangle^2}{m_{e}^2}
+C_{\eta\eta} \langle \eta \rangle^2 + C_{\lambda\lambda} 
\langle \lambda \rangle^2 +C_{m\eta} \langle \eta \rangle \frac{\langle m \rangle}{m_e} 
	\nn
	\eea
	\be
+ C_{m\lambda}
\langle \lambda \rangle \frac{\langle m_{\nu} \rangle}{m_e} 
+C_{\eta\lambda} 
\langle \eta \rangle \langle \lambda \rangle,
	\ee

\be{}
\langle m \rangle = 
|m^{(1)}_{ee}| + e^{i\phi_{2}} |m_{ee}^{(2)}|
+  e^{i\phi_{3}} |m_{ee}^{(3)}|~,
\label{mee}
\ee
	where $m_{ee}^{(i)}\equiv |m_{ee}^{(i)}| \exp{(i \phi_i)}$ 
	($i = 1, 2, 3$)  are  the contributions to $\langle m \rangle$
	from individual mass eigenstates, 
	with  $\phi_i$ denoting relative Majorana phases connected 
	with CP violation, and
	$C_{mm},C_{\eta\eta}, ...$ denote nuclear matrix elements, 
	which can be calculated, (see, e.g. 
\cite{Sta90},  
	for a review see e.g. 
\cite{Tom91,KK60Y,Mut88,Gro89/90,FaesSimc}). 
	Ignoring contributions from right-handed weak currents on the 
	right-hand side of eq.(1), only the first term remains.


\subsection{Allowed Ranges of $\langle m \rangle$ by $\nu$ Oscillation Experiments}

	  The obser\-vable of double beta decay 

\centerline{$\langle m \rangle =
	|m^{(1)}_{ee}| 
		      + e^{i\phi_2} |m^{(2)}_{ee}| 
		      + e^{i\phi_3} |m^{(3)}_{ee}|,
$}

\noindent
	  with $U^{}_{ei}$ 
	  denoting elements of the neutrino mixing matrix, 
	  $m_i$ neutrino mass eigenstates, and $\phi_i$  relative Majorana 
	  CP phases, can be written in terms of oscillation parameters 
\cite{KKPS,KKPS-01} 
\begin{eqnarray}
\label{1}
|m^{(1)}_{ee}| &=& |U^{}_{e1}|^2 m^{}_1,\\
\label{2}
|m^{(2)}_{ee}| &=& |U^{}_{e2}|^2 \sqrt{\Delta m^2_{21} + m^{2}_1},\\
\label{3}
|m^{(3)}_{ee}| &=& |U^{}_{e3}|^2 \sqrt{\Delta m^2_{32} 
				 + \Delta m^2_{21} + m^{2}_1}.
\end{eqnarray}

	The effective mass $\langle m \rangle$ is related with the 
	half-life for $0\nu\beta\beta$ decay via 
$\left(T^{0\nu}_{1/2}\right)^{-1}\sim \langle m_\nu \rangle^2$, 
        and for the limit on  $T^{0\nu}_{1/2}$
	deducible in an experiment we have 
	$T^{0\nu}_{1/2} \sim a \sqrt{\frac{Mt}{\Delta E B}}$.
	Here $a$ is the isotopical abundance of the $\beta\beta$ emitter;
	$M$ is the active detector mass; 
	$t$ is the measuring time; 
	$\Delta E$ is the energy resolution; 
	$B$ is the background count rate. 

%
\begin{figure}[ht]

\centering{
\includegraphics*[scale=0.45]{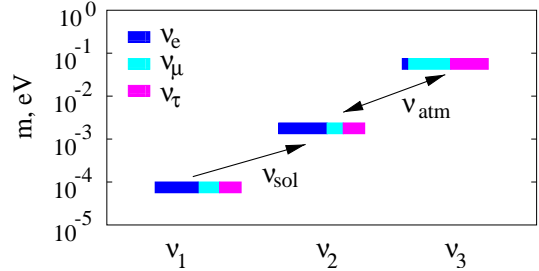}

\vspace{-0.3cm}
\caption[]{
       Neutrino masses and mixings in the scheme with mass hierarchy. 
       Coloured bars correspond to flavor admixtures in the mass 
       eigenstates $\nu_1, \nu_2, \nu_3$. 
       The quantity $\langle m \rangle$
	 is determined by the dark blue bars denoting 
	 the admixture of the electron neutrino $U_{ei}$.
\label{fig:Hierarchi-NuMass}}}
\end{figure}
%

	Neutrino oscillation experiments fix or restrict some of the 
	parameters in 
(1)--(3), e.g. in the case of normal hierarchy solar neutrino 
	  experiments yield 
	  $\Delta m^2_{21}$, 
	  $|U_{e1}|^2 = \cos^2\theta_{\odot}$ 
	  and
	  $|U_{e2}|^2 = \sin^2\theta_{\odot}$. 
	  Atmospheric neutrinos fix  
	  $\Delta m^2_{32}$, 
	  and experiments like CHOOZ, looking for $\nu_e$ 
	  disappearance restrict $|U_{e3}|^2$. 
	  The phases $\phi_i$  and the mass of the lightest neutrino, 
	  $m_1$ are free parameters. 
	Double beta decay can fix the parameter m$_1$ and thus 
	the absolute mass scale.
	  The expectations for 
$\langle m \rangle$ 
	  from oscillation experiments in different neutrino mass scenarios 
	  have been carefully analyzed in  
\cite{KKPS,KKPS-01}. 
	In sections 2.2.1. to 2.2.3. we give some examples.


\vspace{0.3cm}
\noindent
{\it 2.2.1. Hierarchical Spectrum $(m_1 \ll m_2 \ll  m_3)$}
         
\begin{figure}[t]
\centering{\includegraphics*[scale=0.40]{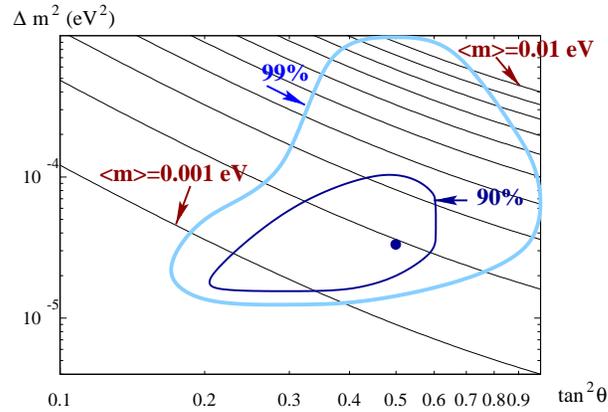}}
\caption[]{
       Double beta decay observable
$\langle m \rangle$
       and oscillation parameters in the case of the MSW 
       large mixing angle solution of the solar neutrino deficit, 
       where the dominant contribution to $\langle m \rangle$ comes 
       from  the  second  state. 
       Shown are  lines of constant $\langle m \rangle$,  
       the lowest  line corresponding to 
       $\langle m_\nu \rangle = 0.001$~eV, 
       the upper line to 0.01~eV. 
       The inner and outer closed line show the regions allowed 
       by present solar neutrino experiments with 
	90\% C.L. and 99\% C.L., 
       respectively. 
       Double beta decay with sufficient sensitivity could check the 
       LMA MSW solution. 
       Complementary information could be obtained from the search for a 
       day-night effect and spectral distortions in future solar 
       neutrino experiments as well as a disappearance signal in KAMLAND 
	[from \cite{KKPS-01}]. 
\label{fig:Dark2}
}
\vspace{.5cm}
\end{figure}

\vspace{0.3cm}
	In hierarchical spectra 
(Fig.~\ref{fig:Hierarchi-NuMass}), 
	motivated by analogies with the quark sector and the simplest 
	  see-saw models, the main contribution comes from 
	  $m_2$ or $m_3$. 
	  For the large mixing angle (LMA) MSW solution which is favored 
	  at present for the solar neutrino problem (see 
\cite{Suz00}), the contribution of $m_2$ becomes dominant in the expression 
       for $\langle m \rangle$, and  
\begin{equation}
\langle m \rangle \simeq m^{(2)}_{ee} 
	= \frac{\tan^2\theta}{1+\tan^2 \theta}\sqrt{\Delta m^2_{\odot}}.
\end{equation}
	In the region allowed at 90\% C.L. by Superkamiokande according to 
\cite{Val01}, 
	the prediction for $\langle m \rangle$ becomes        
\begin{equation}
\langle m \rangle =(1\div 3) \cdot 10^{-3} {\rm eV}.
\end{equation}
	The prediction extends to 
	$\langle m \rangle = 10^{-2}$ eV in the 99\% C.L. range 
(Fig. ~\ref{fig:Dark2}).



\vspace{-0.3cm}
\begin{figure}[h]

\vspace{-0.7cm}
\includegraphics[scale=0.37]
{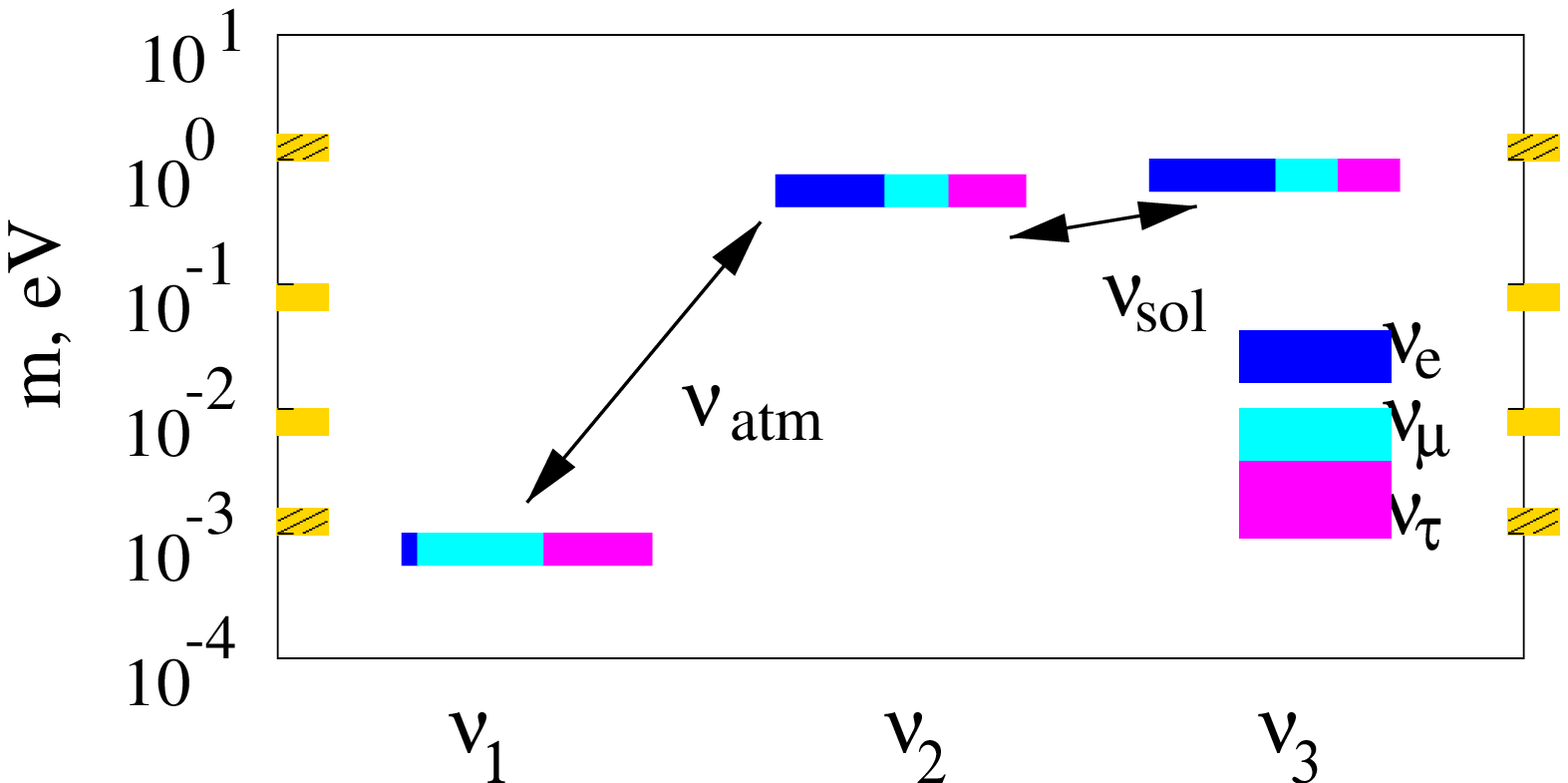}
\includegraphics[scale=0.37]
{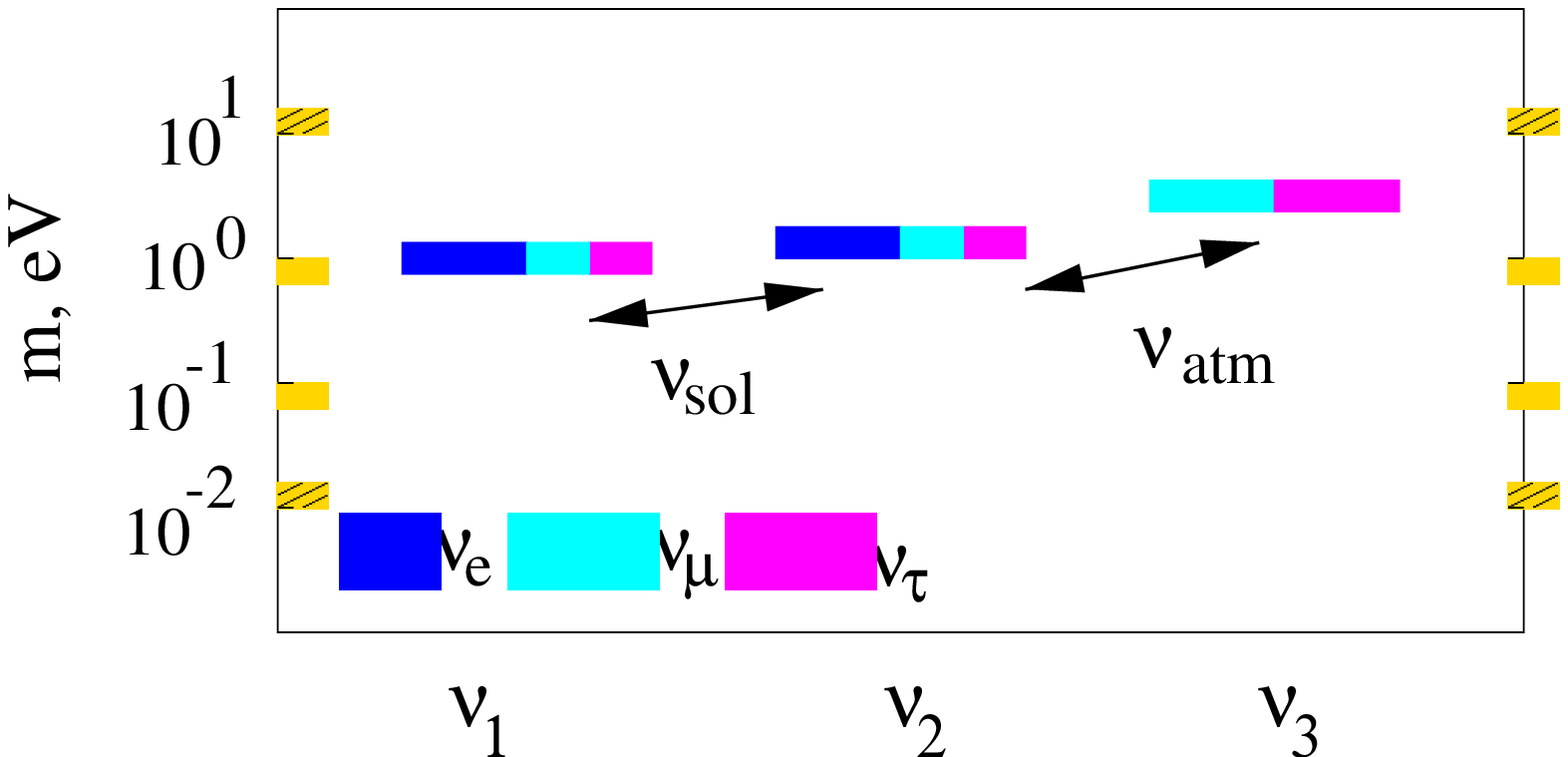}
\caption[]{
       \underline{Left:} Neutrino masses and mixing in the inverse 
	hierarchy scenario. \underline{Right:} Neutrino masses and mixings 
	in the degenerate scheme.
\label{fig:INVERSE-NuMass}
\label{fig:Degener-NuMass}}
\end{figure}


\vspace{0.5cm}
\noindent
{\it 2.2.2. Inverse Hierarchy $(m_3 \approx m_2 \gg  m_1)$}
\vspace{0.3cm}

           In inverse hierarchy scenarios 
(Fig.~\ref{fig:INVERSE-NuMass})   
	the heaviest state with mass $m_3$ is mainly the electron 
	   neutrino, its mass being determined by atmospheric neutrinos, 
	$m_3 \simeq \sqrt{\Delta m^2_{\rm atm}}$.
	   For the LMA MSW solution one finds 
\cite{KKPS-01}
\begin{equation}
	\langle m \rangle 
	= (1\div 7) \cdot 10^{-2} {\rm eV}.
\end{equation}


\vspace{0.3cm}
\noindent
{\it 2.2.3. Degenerate Spectrum $(m_1 \simeq m_2 \simeq m_3 
	\gsim	 0.1~$eV)}
\vspace{0.3cm}

        In degenerate scenarios (fig. \ref{fig:Degener-NuMass}) the 
	contribution of $m_3$ is strongly restricted by CHOOZ.  
	The main contributions come from $m_1$ and $m_2$, depending on 
	their admixture to the electron flavors, which is determined 
	by the solar neutrino solution. We find 
\cite{KKPS-01}
\begin{equation}
m_{\min} < \langle m \rangle < m_1 \qquad 
\mbox{with} \qquad 
\langle m_{\min}\rangle = 
	(\cos^2\theta_{\odot} -\sin^2\theta_{\odot})\, m^{}_1.
\end{equation}

       	This leads for the LMA solution to 
	$\langle m \rangle = (0.25\div 1)\cdot m_1$, 
	 the allowed range corresponding to possible values of 
	 the unknown Majorana CP-phases.

	 After these examples we give a summary of our analysis 
\cite{KKPS,KKPS-01} 
	of the $\langle m \rangle $ allowed by $\nu$ oscillation 
      	experiments for neutrino mass models in the presently 
      	favored scenarios, 
in Fig.~\ref{fig:Jahr00-Sum-difSchemNeutr}. 
	  The size of the bars corresponds to the uncertainty in 
	  mixing angles and the unknown Majorana CP-phases.



\subsection{Status of $\beta\beta$ Research and Evidence for the Neutrinoless Decay Mode}

\begin{figure}[h]
\centering{
\includegraphics*[scale=0.40]{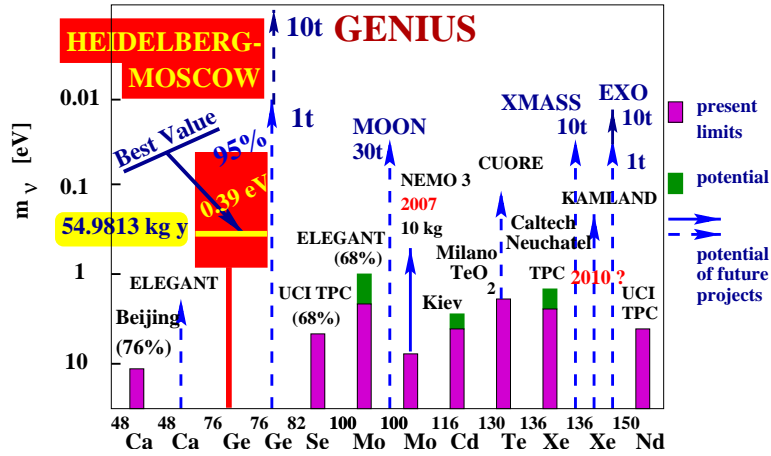}}
\caption[]{
       Present sensitivity, and expectation for the future, 
       of the most promising $\beta\beta$ experiments. 
       Given are limits for $\langle m \rangle $, except 
	for the HEIDELBERG-MOSCOW experiment where the recently 
	observed {\it value}
\cite{KK01-Ev} 
	is given (95$\%$ c.l. range), and the best value.
	Framed parts of the bars: present status; not framed parts: 
       expectation for running experiments; solid and dashed lines: 
       experiments under construction or proposed, respectively. 
       For references see 
\protect\cite{KK60Y,KK-NOON00,KK-NANPino00}.
\label{fig:Now4-gist-mass}}
\end{figure}



	The status of present double beta experiments is shown in 
Fig.~\ref{fig:Now4-gist-mass}	
	and is extensively discussed in 
\cite{KK60Y}.	
	The HEIDELBERG-MOSCOW experiment using the largest source strength 
	of 11 kg of enriched $^{76}$Ge in form of five HP Ge-detectors 
	is running since August 1990 
	in the Gran-Sasso underground laboratory 
\cite{HDM97,KK-StProc00,KK60Y,HDM01,KK01-Ev}, 
	and is since long time the most sensitive one. 
	We communicate here the status of the analysis 
	of November 2001.	

\vspace{0.3cm}
\noindent
{\it 2.3.1. Data from the HEIDELBERG-MOSCOW Experiment}
\vspace{0.3cm}

	We have analysed
\cite{KK01-Ev}
	the data taken in the period August 1990 - May 2000 
	(54.9813\,kg\,y, or 723.44 mol-years 
	and the data of single site events taken in the period 
	November 1995 - May 2000 with our methods of pulse shape 
	analysis (PSA) 
\cite{HelKK00,KKMaj99,Patent-KKHel}, 
	with various statistical methods, in particular with the  
	Bayesian method (see, e.g. 
\cite{Bayes_Method-General}).  
	This method is particularly suited for low
	counting rates, where the data follow 
	a Poisson distribution, that cannot be approximated by a Gaussian.

	This Bayesian procedure reproduces $\gamma$-lines 
	at the position of  known weak lines  
	from the decay of $^{214}{Bi}$ at 2010.7, 2016.7, 2021.8 
	and 2052.9\,keV 
\cite{Tabl-Isot96}. 
	In addition, a line centered at 2039 keV shows up. 
	This is compatible with the Q-value 
\cite{Old-Q-val,New-Q-2001}
	of the double beta decay process. 
	We emphasize, 
	that at this energy no $\gamma$-line is expected 
	according to Monte Carlo simulations of our experimental setup, 
	and to the compilations in 
\cite{Tabl-Isot96}. 
	Therefore, on 
	the Bayesian analysis 
	yields a confidence level  
	for a line to exist at 
	2039.0 keV of 97 $\%$ c.l. (2.2 $\sigma$).  

\vspace{0.3cm}
\noindent
{\it 2.3.2. $\beta\beta$ - Half-Life and Effective Neutrino Mass}
\vspace{0.3cm}

	Under the assumption that the signal at Q$_{\beta\beta}$ is not 
	produced by a background line of at present unknown origin, 
	we can translate the observed number of events into half-lifes 
	for neutrinoless double beta decay. 
	
	In Table 
\ref{Results} 
	we give the values obtained 
	with the Bayesian method. 
	Also given are the effective neutrino mass 
	$\langle m \rangle$ deduced using matrix elements from%
\cite{Sta90}.


\begin{table}[h]

\vspace{-0.3cm}
\caption[]{Half-life for the neutrinoless decay mode 
	and deduced effective neutrino mass 
	from the HEIDELBERG-MOSCOW experiment%
\cite{KK01-Ev}.}
%
\begin{center}
\newcommand{\m}{\hphantom{$-$}}
\renewcommand{\arraystretch}{.8}
\setlength\tabcolsep{6.pt}
\begin{tabular}{c|c|c|c|c}
\hline
\hline
&&&&\\
Significan-	&	Detectors		
&	${\rm T}_{1/2}^{0\nu}	{\rm~ \,y}$	
& $\langle m \rangle $ eV	&	Conf. \\
	ce $[ kg\,y ]$	&&&& level\\
\hline
&&&&\\
	54.9813	
&	1,2,3,4,5
&	$(0.80 - 35.07) \times 10^{25}$	
& (0.08 - 0.54)		
& 	$95\%  ~c.l.$	\\
	54.9813	
&	1,2,3,4,5
&	$(1.04 - 3.46) \times 10^{25}$	
& (0.26 - 0.47)	
& $68\%  ~c.l.$	\\
 	54.9813	
&	1,2,3,4,5
&	$1.61 \times 10^{25}$	
& 0.38 
& Best Value	\\
\hline
 	46.502	
&	1,2,3,5
&	$(0.75 - 18.33) \times 10^{25}$	
& (0.11 - 0.56)
& $95\%  ~c.l.$	\\
	46.502	
&	1,2,3,5
&	$(0.98 - 3.05) \times 10^{25}$	
& (0.28 - 0.49)
& $68\% ~c.l.$	\\
	46.502	
&	1,2,3,5
&	$1.50 \times 10^{25}$	
& 0.39
& Best Value		\\
\hline
	28.053	
&	2,3,5  SSE	
&	$(0.88 - 22.38) \times 10^{25}$	
& (0.10 - 0.51)	
& $90\% ~c.l.$		\\
	28.053	
&	2,3,5  SSE	
&	$(1.07 - 3.69) \times 10^{25}$	
& (0.25 - 0.47)		
& $68\% ~c.l.$		\\
	28.053	
&	2,3,5 SSE
&	 $1.61 \times 10^{25}$	
& 0.38
& Best Value	\\
\hline
\hline
\end{tabular}
\end{center}
\label{Results}
\end{table}


	We derive from the data taken with 46.502\,kg\,y 
	the half-life 
	${\rm T}_{1/2}^{0\nu} = (0.8 - 18.3) \times 10^{25}$ 
	${\rm y}$ (95$\%$ c.l.). 
	The analysis of the other data sets, shown in Table 
\ref{Results},
	and in particular of the single site events data, 
	which play an important role in our  
	conclusion, confirm this result.

	The result obtained is consistent with the limits 
	given earlier by the HEI\-DELBERG-MOSCOW experiment 
\cite{HDM01}. 
	It is also consistent with all other double beta experiments -  
	which still reach less sensitivity. 
	A second Ge-expe\-riment, 
	which has stopped operation in 1999 after 
	reaching a significance of 9\,kg\,y
\cite{DUM-RES-AVIGN-2000}
	yields (if one believes their method of 'visual inspection' 
	in their data analysis) in a conservative analysis a limit of 
	${\rm T}_{1/2}^{0\nu} > 0.55 \times 10^{25}
	{\rm~ y}$  (90\% c.l.). 
	(and not the value given in 
\cite{DUM-RES-AVIGN-2000}).
	The $^{128}{Te}$ geochemical experiment 
\cite{Ber92}
	yields $\langle m_\nu \rangle < 1.1$ eV (68 $\%$ c.l.), 
	the $^{130}{Te}$ cryogenic experiment yields 
	$\langle m_\nu \rangle < 1.8$\,eV 
\cite{Ales00},
	and the CdWO$_4$ experiment $\langle m_\nu \rangle < 2.6$\,eV,  
\cite{Dan2000}, 
	all derived with the matrix elements of 
\cite{Sta90} 
	to make the results comparable to the present value. 

	Concluding we obtain, with about 95$\%$ probability, 
	first evidence for the neutrinoless 
	double beta decay mode. 
	As a consequence, with 95$\%$ confidence, 
	lepton number is not conserved. 
	Further the neutrino is a Majorana particle. 
	The effective mass 
	$\langle m \rangle $ is deduced 
	to be $\langle m \rangle $ 
	= (0.11 - 0.56)\,eV (95$\%$ c.l.), 
	with best value of 0.39\,eV. 
	Allowing conservatively for an uncertainty of the nuclear 
	matrix elements of $\pm$ 50$\%$
	(for detailed discussions of the status 
	of nuclear matrix elements we refer to 
\cite{KK60Y} 
	and references therein like 
\cite{Tom91,Vog2001,FaesSimc}) 
	this range may widen to 
	$\langle m \rangle $ 
	= (0.05 - 0.84)\,eV (95$\%$ c.l.). 

	In this conclusion, it is assumed that contributions 
	to \znbb~ decay from processes other than  
	the exchange of 
	a Majorana neutrino (see, e.g. 
\cite{KK60Y,Moh91}
	and references therein) are negligible.

\subsection{\boldmath Future of $\beta\beta$ Experiments}

	To improve the present sensitivity for the effective neutrino mass 
	considerably, 
	and to fix this quantity more accurately,  
		   requires new experimental approaches, 
		   as discussed extensively in%
\cite{KK60Y,KK-NANPino00,KK-NOW00,KK-Bey97,GEN-prop,KK-Neutr98}. 
	Some of them are indicated in Figs. 
\ref{fig:Now4-gist-mass},\ref{fig:Jahr00-Sum-difSchemNeutr}.

      It has been pointed out earlier (see e.g.%
\cite{KK60Y,KK-NANPino00,KK-NOW00,KK-Bey97,KK-Neutr98,KK-NOW00}), 
	that of present generation experiments probably no one has a 
	potential to probe   
$\langle m\rangle$
     	below (and perhaps even down to) 
	the present HEIDELBERG-MOSCOW level (see 
Fig.~\ref{fig:Now4-gist-mass}).

	 The Milano cryogenic experiment using TeO$_2$ bolometers 
	 improved their values for the 
$\langle m \rangle$ 
	 from $\beta\beta$ decay of $^{130}$Te, from 5.3 eV in 1994%
\cite{Ales94} 
	to 1.8 eV in 2000%
\cite{Ales00}.
	NEMO-III, originally aiming at a sensitivity of 0.1 eV, 
	reduced their goals recently to $0.3\div0.7$~eV (see%
\cite{NEMO-Neutr00}
	) (which is more consistent with estimates given by%
\cite{Tret95}
	), to be reached in 6 years from starting of running, 
       foreseen for the year 2002.


\vspace{.3cm}
\noindent
{\it 2.4.1. GENIUS and other Proposed Future Double Beta Experiments}

\vspace{.3cm}
	With the era of the HEIDELBERG-MOSCOW experiment 
	the time of the small smart 
	experiments is over.

	     To reach 
	significantly larger sensitivity, $\beta\beta$ 
	     experiments have to become large. 
	     On the other hand source strengths of up to 10 tons of 
	     enriched material touch the world production limits. 
	This means that the background has to be reduced by the order a 
	     factor of 1000 and more compared to that 
	     of the HEIDELBERG-MOSCOW experiment.


\begin{table*}[b]
\caption{Some key numbers of future double beta decay experiments (and of 
	the {\sf HEIDELBERG-MOSCOW} experiment). Explanations: 
	${\nabla}$ - assuming the background of the present pilot project. 
	$\ast\ast$ - with matrix element from  
\protect\cite{Sta90}, 
\protect\cite{Tom91}, 
\protect\cite{Hax84}, 
\protect\cite{WuStKKChTs91}, 
\protect\cite{WuStKuKK92} (see Table II in 
\protect\cite{HM99})  
	and assuming an uncertainty of $\pm$50$\%$ 
	of the nuclear matrix element.
	${\triangle}$ - this case shown 
	to demonstrate {\bf the ultimate limit} of such experiments. 
	For details see 
\protect\cite{KK60Y}.}
\label{table:1}
\newcommand{\m}{\hphantom{$-$}}
\newcommand{\cc}[1]{\multicolumn{1}{c}{#1}}
\renewcommand{\tabcolsep}{.04 pc} 
\renewcommand{\arraystretch}{.2} 
{\footnotesize
{
\begin{center}  
\begin{tabular}[!h]{|c|c|c|c|c|c|c|c|}
\hline
\hline
 &  &  &  & Assumed &  &  & \\
 &  &  &  & backgr. & $Running$ & Results & \\
$\beta\beta$-- & & & Mass & $\dag$ events/ & 	 & limit for & 
${<}m_{\nu}{>}$ \\
$Isoto-$ & $Name$ & $Status$ & $(ton-$ & kg y keV, & Time  
& $0\nu\beta\beta$ & \\
pe & & & $nes)$ & $\ddag$ events/kg & 	 & half-life & ( eV )\\ 
& & & & y FWHM,  & (tonn. & (years) & \\
& & & & $\ast$ events & years) &  & \\
& & &  & /yFWHM &  &  & \\
\hline
\hline
 &  &  &  &  &  &  & \\
~${\bf ^{76}{Ge}}$ & {\bf HEIDEL-} & {\bf run-}  & 0.011 & $\dag$ 0.06 
& {\bf 54.98} & ${\bf (0.8-18.3)}$ & {\bf (0.05-0.84)}\\
 & {\bf BERG}  & {\bf ning}  &  (enri-  &  &  {\bf kgy} 
&  ${\bf x~{10}^{25} y}$ 	& 	{\bf eV}$)^{\ast\ast}$\\
& {\bf MOSCOW} & {\bf since} & ched) & $\ddag$ 0.24  &  
& {\bf 95$\%$ c.l.}	& {\bf 95$\%$ c.l.} \\
& {\bf \cite{KK01-Ev,HDM01,KK-SprTracts00}} & {\bf 1990} &  & $\ast$ 2 & & 
{\bf NOW !!}  &  {\bf NOW !!}\\
\hline
\hline
\hline
 &  &  &  &  &  &  & \\
${\bf ^{100}{Mo}}$ & {\sf NEMO III} & {\it under} & $\sim$0.01 & $\dag$ 
{\bf 0.0005} &  &  &\\
 & {\tt \cite{NEMO-Neutr00}}& {\it constr.} & (enri- & $\ddag$ 0.2  & 50 & 
${10}^{24}$ & 0.3-0.7\\
 &  & {\it end2001?} & -ched) &  $\ast$ 2 &kg y  &  &\\
\hline
\hline
&  &  &  &  &  &  & \\
${\bf ^{130}{Te}}$ & ${\sf CUORE}^{\nabla}$ & {\it idea} & 0.75 & $\dag$ 0.5 
& 5 & $9\cdot{10}^{24}$ & 0.2-0.5\\
 & {\tt \cite{CUORE-LeptBar98}}& {\it since1998} &(nat.)  
& $\ddag$ 4.5/$\ast$ 1000 &  & & \\ 
\hline
&  &  &  &  &  &  &  \\
${\bf ^{130}{Te}}$ & {\sf CUORE}  &  {\it idea} & 0.75   & $\dag$ 0.005 & 5 
& $9\cdot{10}^{25}$ & 0.07-0.2\\
&  {\tt \cite{CUORE-LeptBar98,Fior-Neutr00}} & {\it since1998}   & (nat.)  
&  $\ddag$ 0.045/ $\ast$ 45 &  & &\\
\hline
&  &  &  &  &  &  & \\
${\bf ^{100}{Mo}}$ & {\sf MOON} & {\it idea} & 10 (enr.) & ? & 30 & ? &\\
 & {\tt \cite{Ej00,LowNu2}} & {\it since1999} &  100(nat.) & & 300 & &0.03 \\
\hline
&  &  &  &  &  &  & \\
${\bf ^{116}{Cd}}$ & {\sf CAMEOII} & {\it idea}  & 0.65 & * 3. & 5-8  
& ${10}^{26}$ & 0.06 \\
& {\sf CAMEOIII}{\sf \cite{Bell00}} & {\it since2000 } & 1(enr.) & ? & 5-8 
&  ${10}^{27}$ & 0.02 \\
\hline
&  &  &  &  &  &  & \\
${\bf ^{136}{Xe}}$ & {\sf EXO} & Proposal& 1 & $\ast$ 0.4 & 5 & 
$8.3\cdot{10}^{26}$ & 0.05-0.14\\
&  & since &  &  &  &  & \\
  & {\tt \cite{EXO00,EXO-LowNu2}} & 1999 & 10 & $\ast$ 0.6 & 10 & 
$1.3\cdot{10}^{28}$ & 0.01-0.04\\
\hline 
\hline
\hline
\hline
&  &  &  &  &  &  &  \\
~${\bf ^{76}{Ge}}$ & {\bf GENIUS} & {\it under} & 11 kg & 
$\dag$ ${\bf 6\cdot{10}^{-3}}$& 3 
& {\bf ${\bf 1.6\cdot{10}^{26}}$} & {\bf 0.15} \\
& {\bf - TF} & {\it constr.} & (enr.) &  &  &  &   \\
&  {\bf \cite{KK-GeTF-MPI,GenTF-0012022}}&  {\it end 2002} &  & &  &  &  \\
\hline
&  &  &  &  &  &  &  \\
~${\bf ^{76}{Ge}}$ & {\bf GENIUS} & Pro- & 1  & $\dag$ 
${\bf 0.04\cdot{10}^{-3}}$ & 1 & ${\bf 5.8\cdot{10}^{27}}$ & 
{\bf 0.02-0.05} \\
 & {\tt \cite{KK-Bey97,GEN-prop,KK-Neutr98}}  & posal &(enr.)  
& $\ddag$ ${\bf 0.15\cdot{10}^{-3}}$ & & & \\
&  & since &  & $\ast$ {\bf 0.15} &  &  &  \\
&  & 1997 & 1 & ${\bf \ast~ 1.5}$ & 10 & ${\bf 2\cdot{10}^{28}}$  & 
{\bf 0.01-0.028} \\
\hline
&  &  &  &  &  &  &  \\
~${\bf ^{76}{Ge}}$ & {\bf GENIUS} & Pro- & 10 
& $\ddag$ ${\bf 0.15\cdot{10}^{-3}}$ & 10 &
${\bf 6\cdot{10}^{28}}$ & {\bf 0.006 -}\\
&  {\tt \cite{KK-Bey97,GEN-prop,KK-Neutr98}} &  posal 
&  &  &  &  &  {\bf 0.016}\\
 &   &  since &(enr.) & ${\bf 0^{\triangle}}$ & 10 & 
${\bf 5.7\cdot{10}^{29}}$ & {\bf 0.002 -}\\
&  &  1997 &  &  &  &  &  {\bf 0.0056}\\ 
\hline 
\hline
\end{tabular}\\[2pt]
\end{center}}}
\label{List-exp}
\end{table*}


	Table 
\ref{List-exp} 
	lists some key numbers for GENIUS%
\cite{KK-Bey97,GEN-prop,KK-NOW00}, 
	     which was the 
	     first proposal for a third generation double beta experiment, 
	and of some other 
	     proposals made after the GENIUS proposal. 
	     The potential of some of them is shown also in 
Fig.~\ref{fig:Jahr00-Sum-difSchemNeutr}, and 
	it is seen that not all of them will lead to large 
	improvements in sensitivity.
	Among the latter is also the recently presented MAJORANA project%
\cite{MAJOR-WIPP00}, 
	which does not really apply a striking new strategy 
	for background reduction, 
	particularly also after it was found that the projected  
	segmentation of detectors may not work (see Table 3).

\begin{table*}[h]
\begin{center}
\caption{\label{New-Proj}Some of the new projects under discussion for future double beta 
	decay experiments (see ref.%
\protect\cite{KK60Y}).}
\label{tableA}
\newcommand{\m}{\hphantom{$-$}}
\newcommand{\cc}[1]{\multicolumn{1}{c}{#1}}
\renewcommand{\tabcolsep}{.19 pc} 
\renewcommand{\arraystretch}{.05} 
\begin{tabular}[!h]{|c|c|c|c|c|}
\hline
\hline
\multicolumn{5}{|c|}{}\\
\multicolumn{5}{|c|}{NEW~~~  PROJECTS}\\
\multicolumn{5}{|c|}{}\\
\hline
 &  &  &  & \\
 & BACKGROUND & MASS & POTENTIAL & POTENTIAL \\
 &  &  &  & \\
 & REDUCTION & INCREASE & FOR DARK & FOR SOLAR\\
 &  &  & MATTER & ${\nu}^{'}$ s\\
\hline
&  &  &  & \\
 {\bf GENIUS} 	& {\bf +} 	& {\bf +} 	
& {\bf +} & {\bf + ${}^{*)}$	}\\
\hline
&  &  &  & \\
 {\sf CUORE}  	& {\bf (+)} 	& {\bf + } 	
& {\bf $-$} & {\bf $-$	}	\\
\hline
&  &  &  & \\
 {\sf MOON}  		& {\bf (+) }	& {\bf + } 	
& {\bf $-$} & {\bf +}		\\
\hline
&  &  &  & \\
 {\bf EXO} 		&  {\bf +} 	& {\bf +}  	
& {\bf $-$} & {\bf $-$	}	\\
\hline
&  &  &  & \\
 {\sf MAJORANA}	& {\bf $-$ }	& {\bf + } 	
& {\bf $-$} & {\bf $-$	}	\\
\hline
\multicolumn{5}{|c|}{}\\
\multicolumn{5}{|c|}{\sf *) real time measurement of pp neutrinos 
	with threshold of 20 keV (!!)}\\
\multicolumn{5}{|c|}{}\\
\hline
\hline
\end{tabular}\\[1pt]
\end{center}
\end{table*}


	For more recent information on XMASS, EXO, MOON experiments see 
	the contributions of 
	Y. Suzuki, G. Gratta and H. Ejiri in Ref. 
\cite{LowNu2}.
	     The CAMEO project 
\cite{Bell00}
	in its now propagated variant GEM  
	is nothing then 
	a variant of GENIUS (see below) put into the BOREXINO tank, 
	at some later time. 	
	CUORE  
\cite{CUORE-LeptBar98} 
	has, with the complexity of cryogenic techniques, 
	   still to overcome serious 
	   problems of background to enter 
	   into interesting regions of
$\langle m_\nu\rangle$.
	 EXO  
\cite{EXO00} 
	needs still very extensive research and development to probe 
	 the applicability of the proposed detection method.
	In particular if it would be confirmed that tracks will 
	be too short to be identified, it would act essentially 
	only as a highly complicated calorimeter.
	 In the GENIUS project a reduction by a factor of more than 1000 
	 down to a background level of 0.1 events/tonne y keV 
	 in the range of $0\nu\beta\beta$ decay is 
	planned to be reached by removing all 
	 material close to the detectors, and by using naked Germanium 
	 detectors in a large tank of liquid nitrogen. 
	 It has been shown that the detectors show excellent 
	 performance under such conditions  
\cite{GEN-prop,KK-J-PhysG98}.
	For technical questions and extensive Monte Carlo simulations of 
	the GENIUS project for its application in double beta decay 
	we refer to  
\cite{GEN-prop,KK-J-PhysG98}.

\vspace{0.3cm}
\noindent
{\it 2.4.2. GENIUS and Other Beyond Standard Model Physics}
\vspace{0.3cm}

	GENIUS will allow besides the 
	large increase in sensitivity for double beta decay 
	described above, the access to a broad range 
		   of other beyond SM physics topics in the multi-TeV range. 
	Already now $\beta\beta$ decay probes the TeV scale on which new 
	physics should manifest itself (see, e.g.%
\cite{KK60Y,KK-Bey97,KK-LeptBar98,KK-SprTracts00,KK-InJModPh98}). 
	  Basing to a large extent on the theoretical work of the 
	  Heidelberg group in the last six years, the 
	  HEIDELBERG-MOSCOW experiment yields results for SUSY models 
	  (R-parity breaking, neutrino mass), leptoquarks 
	  (leptoquarks-Higgs coupling), compositeness, right-handed $W$ mass, 
	  nonconservation of Lorentz invariance and 
	  equivalence principle, mass of a heavy left or 
	  righthanded neutrino, competitive to corresponding results 
	  from high-energy accelerators like TEVATRON and HERA 
	(for details see 
\cite{KK60Y,KK-LeptBar98,KK-SprTracts00}).


\vspace{0.3cm}
\noindent
{\it 2.4.3. GENIUS-Test Facility}
\vspace{0.3cm}

		     Construction of a test facility for 
		     GENIUS -- GENIUS-TF -- 
		     consisting of $\sim$~40\,kg of HP Ge 
		     detectors suspended in a liquid nitrogen 
		     box has been started. 
		     Up to summer of  2001, six detectors each 
		     of $\sim$~2.5\,kg and with a threshold as low 
		     as $\sim$~500\,eV have been produced.

         Besides test of various parameters of the GENIUS project, 
	 the test facility would allow, with the projected background 
	 of 2--4\,events/(kg\,y\,keV) in the low-energy range, 
	to probe the DAMA 
	 evidence for dark matter by the seasonal modulation signature, 
	(see%
\cite{KK-GeTF-MPI,GenTF-0012022},
	and section 3).




\section{Dark Matter}

\subsection{SUSY Expectations for Cold Dark Matter}

	Direct search for WIMPs can be done

	(a) by looking for the recoil nuclei 
	in WIMP- nucleus elastic scattering. 
	The signal could be ionisation, phonons or light produced 
	by the recoiling nucleus. 
	The typical recoil energy is a few 100\,eV/GeV WIMP mass.

	(b) by looking for the modulation of the WIMP  
	signal resulting from the seasonal variation 
	of the earth's velocity against the WIMP 'wind'.

	The expectation for neutralino elastic scattering 
	cross sections and masses have been extensively 
	analysed in many variants of SUSY models.


\begin{figure}[h]
\includegraphics*[width=65mm, height=55mm]
{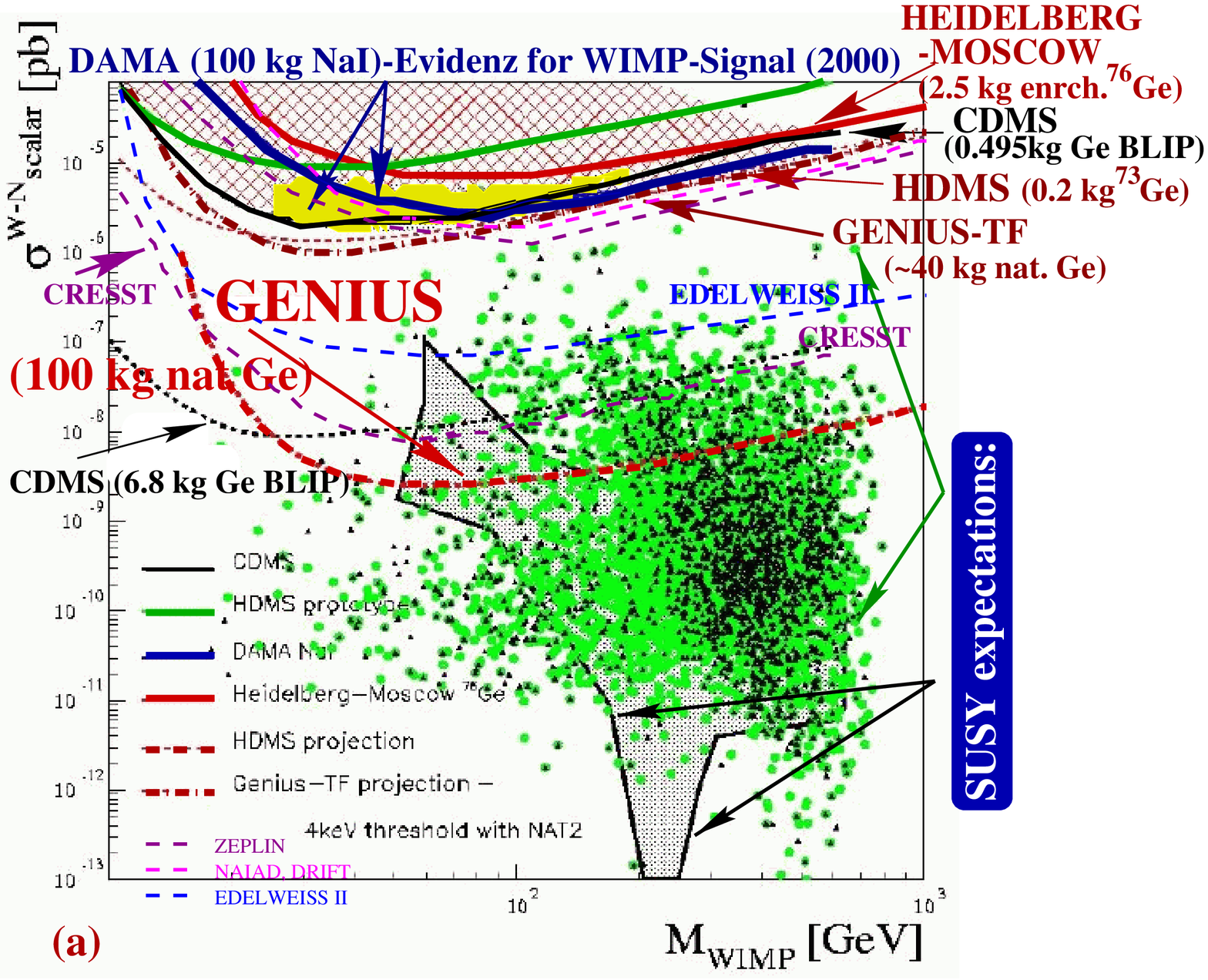}
\hspace{-.5cm}
\includegraphics*[width=57mm, height=50mm]
{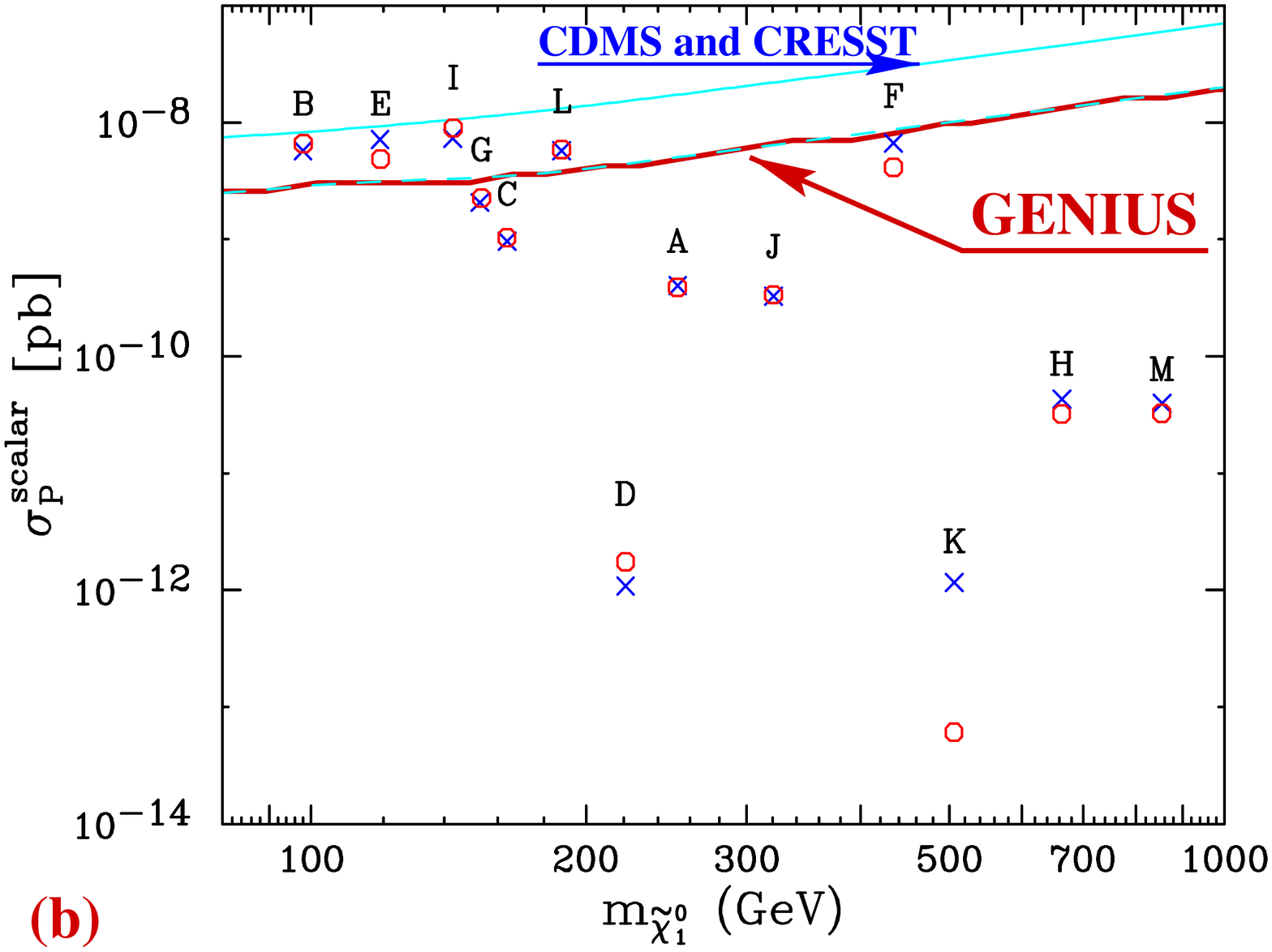}

\caption[]{
       (a): WIMP-nucleon cross section limits in pb 
	for scalar interactions as 
       function of the WIMP mass in GeV. 
       Shown are contour lines of present experimental limits (solid lines) 
       and of projected experiments (dashed lines). 
       Also shown is the region of evidence published by DAMA. 
       The theoretical expectations from the MSSM 
	are shown by two scatter plots, 
	- for accelerating and for non-accelerating Universe (from%
\cite{BedKK00_01}), 
	and from the SUGRA by the grey region (from%
\cite{EllOliv-DM00}). 
	{\em Only}~ GENIUS will be able to probe the shown range 
       also by the signature from seasonal modulations. 
	(b): WIMP- proton elastic scattering cross sections 
	according to various MSUGRA models (see text). From%
\protect\cite{Ell-New-01-11}.
\label{fig:Bedn-Wp2000}}
\end{figure}

\begin{figure}[htb]

\vspace{-0.5cm}
\begin{center}
\includegraphics*[width=75mm, height=75mm]
{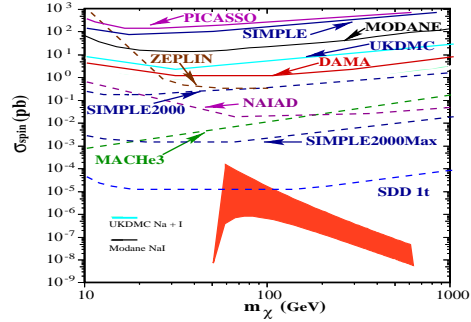}
\end{center}

\vspace{-0.9cm}
\caption{The same in Figs.%
\ref{fig:Bedn-Wp2000}a, 
	but for spin-dependent interaction. 
	The grey area corresponds to the MSUGRA expectations%
\protect\cite{EllOliv-DM00}. 
	Full lines denote present experimental results, 
	dashed lines expectations of future experiments.}
\label{fig:Ell-New}
\end{figure}

	Figs.%
\ref{fig:Bedn-Wp2000},\ref{fig:Ell-New}
	represent the present situation. 
	The SUSY predictions in Figs.%
\ref{fig:Bedn-Wp2000}a,\ref{fig:Ell-New}
	are from the MSSM with relaxed unification conditions%
\cite{BedKK00_01} 
	and the MSUGRA model%
\cite{EllOliv-DM00}.
	Fig.%
\ref{fig:Bedn-Wp2000}b
	shows the result of a study 'at Post-LEP Benchmark points'
	based again on the MSUGRA%
\cite{Ell-New-01-11}. 
	Present experiments only just touch the border 
	of the area predicted by the MSSM. 
	The experimental DAMA evidence for dark matter 
	lies in an area, in which MSUGRA models do not expect dark matter. 
	They would require beyond GUT physics in this frame%
\cite{Arn-priv01}.

\begin{figure}[h]
\begin{center}
\epsfig{file=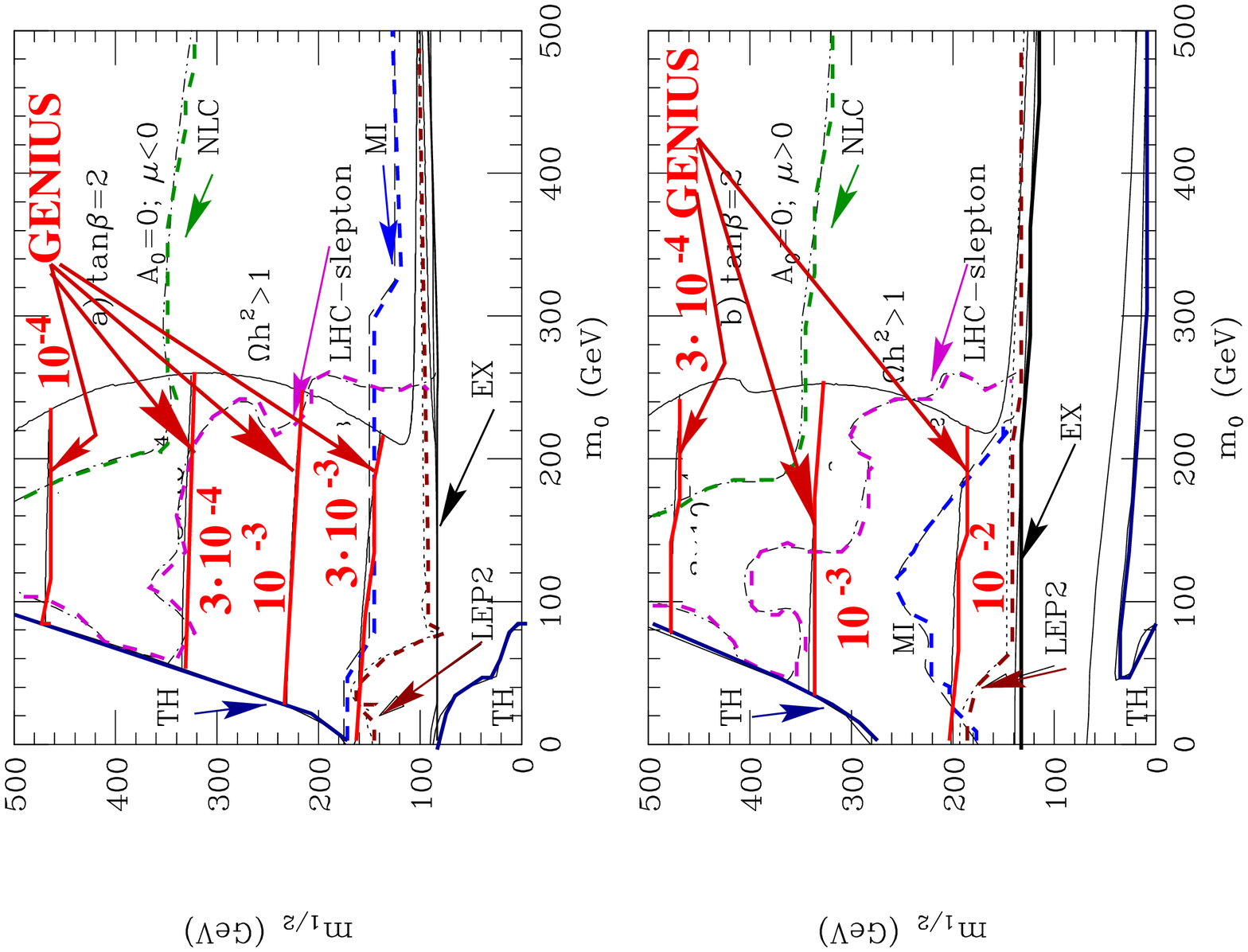,scale=0.35,angle=-90}
\end{center}

\vspace{-0.5cm}
\caption[]{
	Direct detection rate R in events/kg\,day in a $^{73}{Ge}$ detector,  
	and SUSY reach contours for LEP2, Tevatron (MI), 
	LHC slepton signal and NLC. TH - excluded 
	by theoretical considerations. EX - excluded by collider 
	searches for SUSY particles (from%
\cite{Brhlik-Dark98}).
\label{fig:Brhlik}}
\end{figure}

\begin{figure}[h]
\begin{center}
\epsfig{file=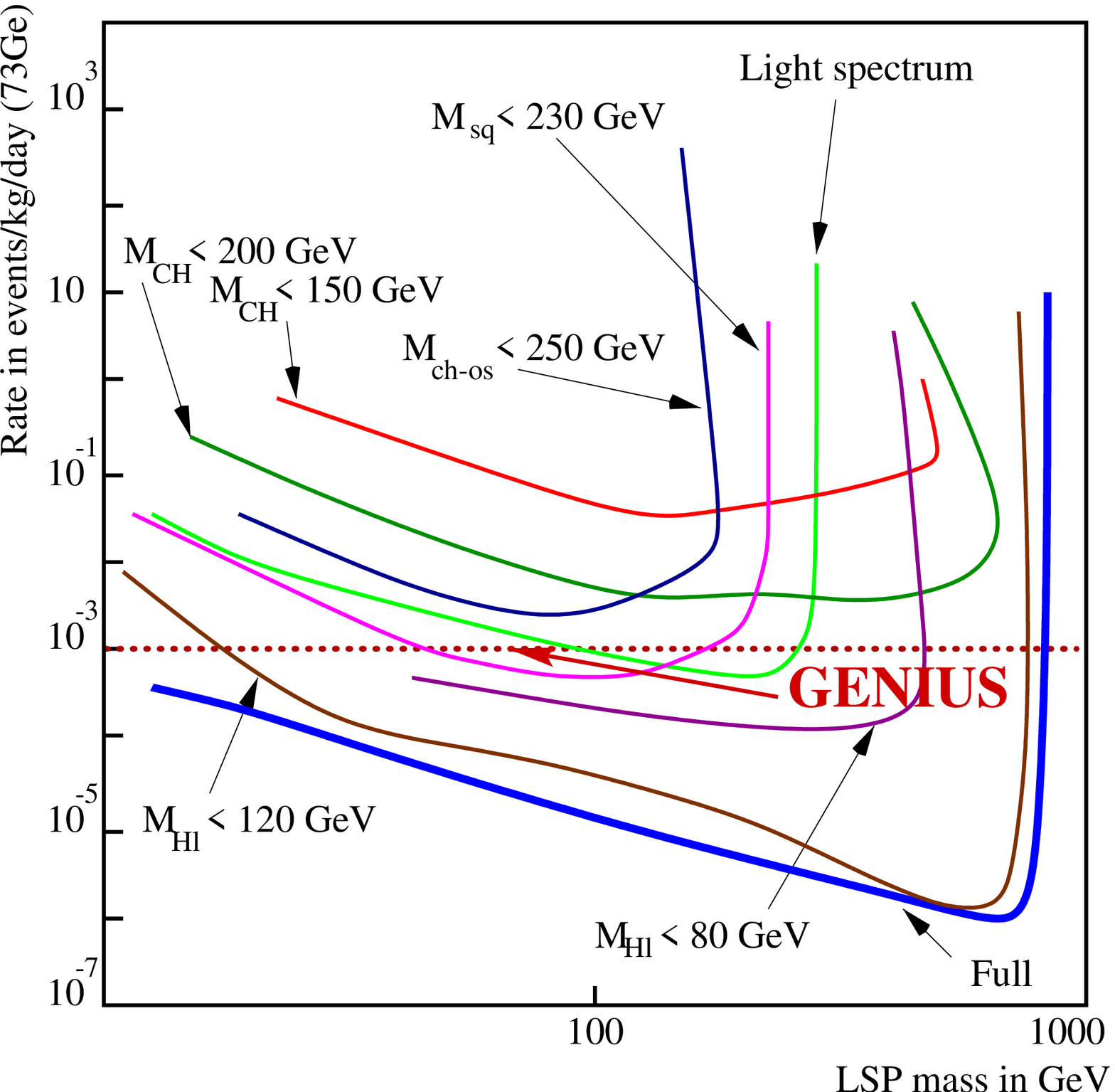,scale=0.3,}
\end{center}

\vspace{-0.2cm}
\caption[]{
	Different lower bounds for the total event rate 
	in $^{73}{Ge}$ (events/day\,kg) versus mass of the LSP (GeV). 
	Here M$_{sq, CH, Hl}$ denote masses of the squark, 
	the charged Higgs boson, and the light neutral 
	CP-even HIggs boson, respectively. 
	Heavy chargino mass is denoted as M$_{ch-os}$. 
	``Full'' corresponds to the lower bound obtained 
	from main (unconstrained) scan, and ``light spectrum'' 
	denotes the lower bound for R, which is obtained with 
	all sfermion masses lighter then about 300\,GeV. 
	The horizontal dotted line represents expected 
	sensitivity for the direct dark matter detection 
	with GENIUS (from%
\cite{BedKK00_01}).
\label{fig:Bedn-constr}}
\end{figure}

	It is clear from extensive theoretical work that high-sensitivity 
	dark matter experiments can yield an important contribution 
	to SUSY search. Fig.%
\ref{fig:Brhlik}
	(from%
\cite{Brhlik-Dark98})
	shows, in the MSUGRA model, the SUSY reach contours 
	for different accelerators (LEP2, Tevatron, LHC, NLC) 
	together with direct detection rates in a $^{73}{Ge}$ detector. 
	It is visible that a detector of GENIUS-sensitivity operates 
	in SUSY search on the level of LHC and NLC. Fig.%
\ref{fig:Bedn-constr} 
	(from%
\cite{BedKK00_01}), 
	shows another study in the MSSM with relaxed unification 
	conditions. Non-observation of Dark Matter with GENIUS 
	would exclude a 'light' SUSY spectrum 
	(all sfermion masses lighter than 300 - 400\,GeV) 
	and any possibility for a light Higgs sector in the MSSM.

	If classifying the SUGRA models 
	into more g$_\mu$-2-friendly (I,L,B,G,C,J) and less 
	g$_\mu$-2-friendly models, according to%
\cite{Ell-New-01-11},
	the former ones have good prospects 
	to be detectable by LHC and/or a 1\,TeV collider. 
	GENIUS could check not only the larger part of these ones, 
	but in addition two of the less 
	g$_\mu$-2-friendly models (E and F), 
	which will be difficult to be probed by future colliders (see Fig.%
\ref{fig:Bedn-Wp2000}b).
	This demonstrates nicely the complementarity 
	of collider and underground research.

\begin{figure}[htb]
%
\begin{center}
\includegraphics*[scale=0.45]{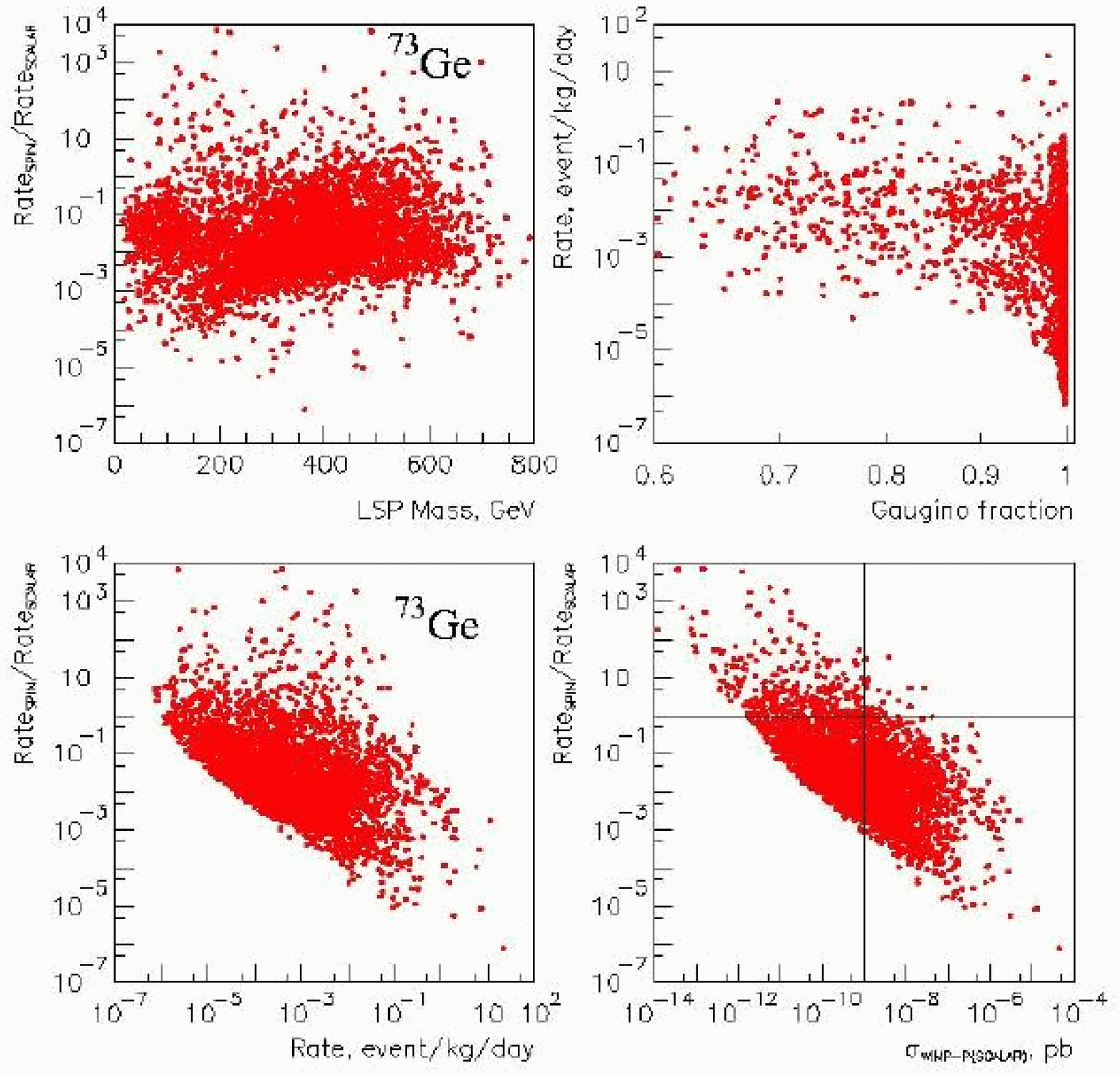} 
\end{center}

\vspace{-0.3cm}
\caption{Ratio of spin-dependent event rate to spin-independent 
	event rate in $^{73}{Ge}$ as function 
	of LSP mass (upper left), of total 
	(spin-dependent plus spin-independent) event rate 
	(lower left part), 
	and of scalar cross section of neutralino-proton interaction 
	(lower right) obtained with 
	0.1$<\Omega_\chi {h_0}^2<$0.3. 
	The vertical line gives the expected sensitivity of GENIUS. 
	In the region above the horizontal line the spin 
	contribution dominates. 
	The total event rate versus gaugino fraction of LSP 
	is also given (upper right). From%
\protect\cite{BedKK00_01}.}
\label{fig:Bedn-Sp}
\end{figure}

	It might be mentioned that in case of g$_\mu$-2 - unfriendly 
	models, i.e. those with very low cross sections in Figs.%
\ref{fig:Bedn-Wp2000}b,
	it might be required to turn from spin-zero targets 
	and looking for spin-independent interaction, 
	which usually for not too light nuclei gives 
	the largest cross sections, to spin- non-zero target nuclei 
	and spin-dependent interaction%
\cite{BedKK00_01}.
	It has been shown recently%
\cite{BedKK00_01}
	that if spin-zero experiments with sensitivities 
	of 10$^{-5}$-10$^{-6}$\,events/kg\,day will fail to detect 
	a dark matter signal, an experiment with nonzero spin target 
	and higher
 sensitivity will be able to detect dark matter 
	{\it only} due to the spin neutralino-quark interaction (see Fig.%
\ref{fig:Bedn-Sp}).

\subsection{Cold Dark Matter - Present and Future Experiments}

	Summarizing the present experimental status, present 
	and also future projects can be categorized in two classes:
	
	1. Sensitivity (or sensitivity goal) 'just for' confirmation of DAMA.

	2. Sensitivity to enter deeply into the range of SUSY predictions.

	Only very few experiments may become candidates 
	for category 2 in a foreseeable future (see Figs.%
\ref{fig:Bedn-Wp2000},\ref{fig:Ell-New}),
	and as far as at present visible, of those 
	only GENIUS will have the chance to search for modulation, 
	i.e. to check, like DAMA, positive evidence for a dark matter signal.

\begin{figure}[h]
\centering{
\includegraphics*[scale=0.3]
{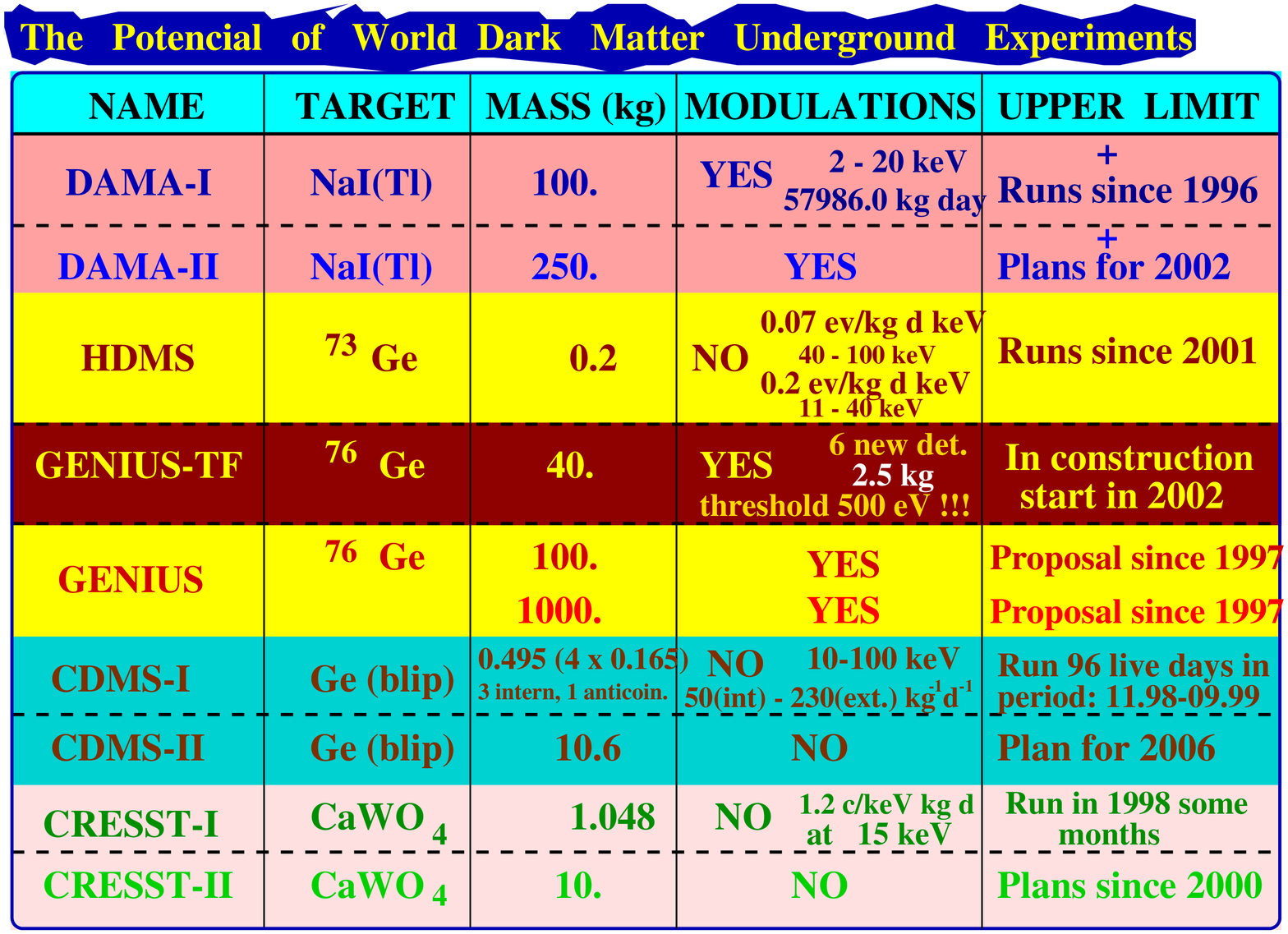}}

\caption[]{
	Status of some running and future Dark Matter experiments.
 \label{fig:table-Status}}
\end{figure}

\begin{figure}[h]
\centering{
\includegraphics*[scale=0.3, angle=-90]
{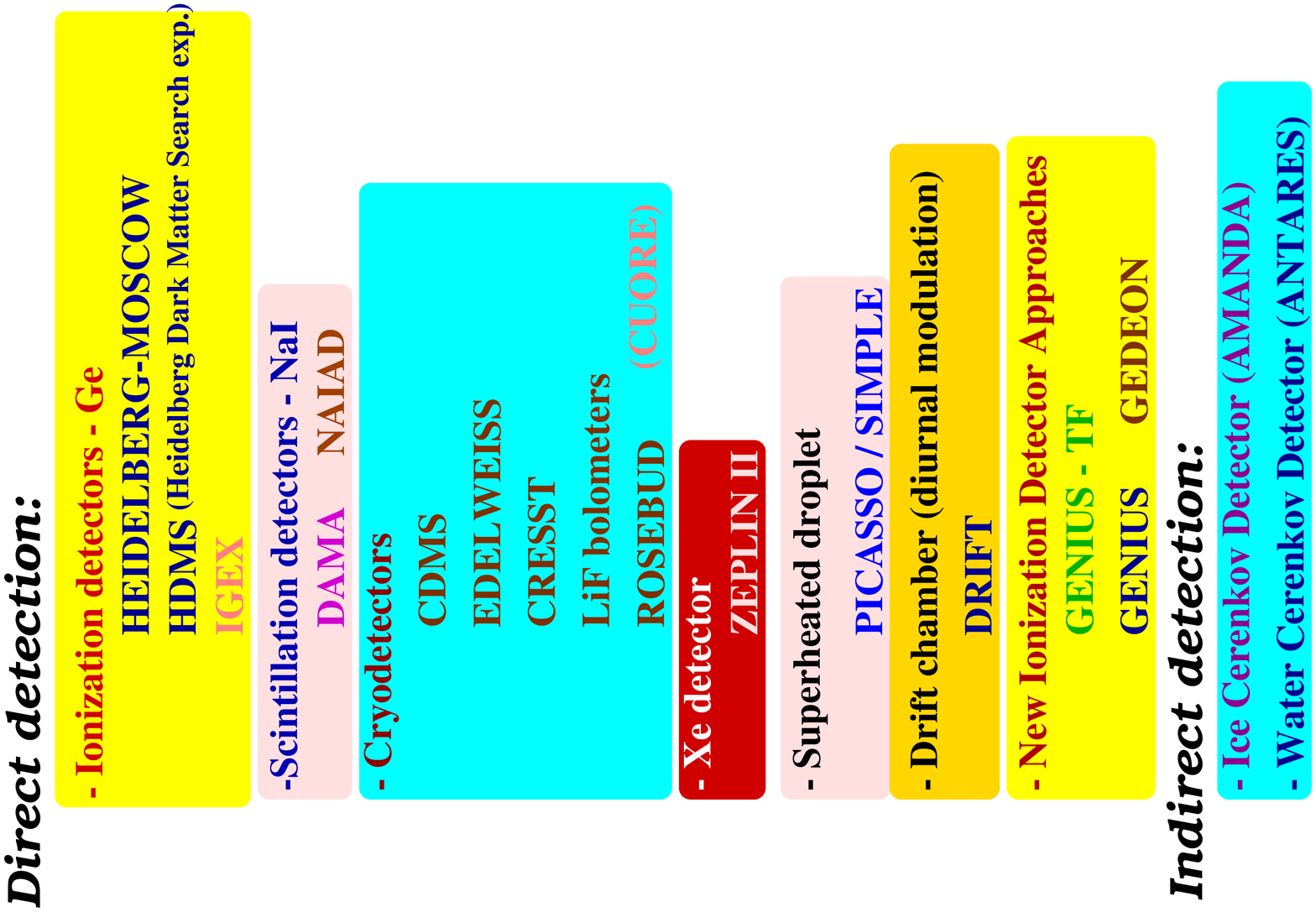}}

\caption[]{
	List of some of the main present and future dark matter experiments.
 \label{fig:table-List}}
\end{figure}

	Figs.%
\ref{fig:table-Status},\ref{fig:table-List} 
	give an overview of present and future experiments. 
	The at present most sensitive experiments DAMA, CDMS 
	(and Edelweiss) are claimed%
\cite{Gait-Nanp2000}
	not to be fully consistent, 
	although CDMS can at present not exclude the full DAMA 
	evidence region%
\cite{DAMA01,Gait-Nanp2000}.
	Some problems in the data analysis 
	of CDMS have been revised recently%
\cite{sad-Taup01}.

	One of the main problem of the cryodetectors 
	is to obtain good numbers of background in the {\it raw} data, 
	i.e. of the starting values for the rejection procedure. 
	This is also the reason, why simple cryogenic 
	calorimeters such as the CRESST I%
\cite{Ram99,CRESST01}
	and the Milano%
\cite{Ales00,CUORE98}
	bolometers are not fully competitive.

\begin{figure}[h]
\centering{
\includegraphics*[scale=0.45]
{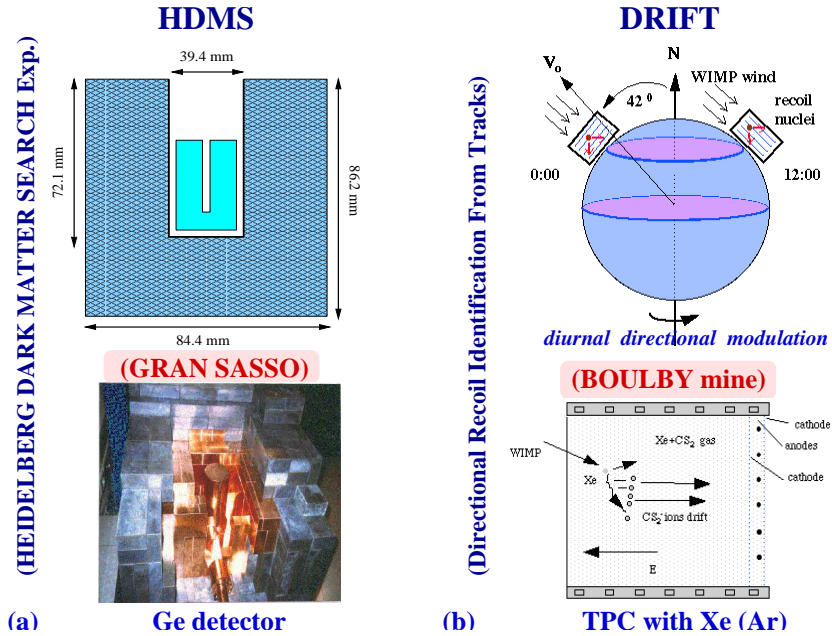}}

\caption[]{
	HDMS and DRIFT experiments.
 \label{fig:HDMS-DRIFT}}
\end{figure}

	The at present most sensitive experiments using raw data 
	are the HEIDELBERG-MOSCOW experiment%
\cite{HM98}
	and the HEIDELBERG Dark Matter Search Experiment (HDMS)%
\cite{KK-dark00,KK-PRD63-00}.
	HDMS uses a special configuration of Ge detectors, 
	with an enriched $^{73}{Ge}$ detector embedded into a 
	natural Ge-detector (Fig.%
\ref{fig:HDMS-DRIFT}a).
	Results of the pilote project are given in%
\cite{KK-dark00,KK-PRD63-00}. 
	The final configuration with the enriched $^{73}{Ge}$ 
	inner detector is in operation since February 2001.

	The experimental future is also illustrated in Figs.%
\ref{fig:Bedn-Wp2000},\ref{fig:Ell-New}. 
	A useful overview is given in%
\cite{DRIFT01}, 
	and for Ge detectors in%
\cite{MORALES}.
	For an earlier review see%
\cite{KKRam98}.

	The cryogenic projects are CDMS II, CREST II, Edelweiss II. 
	In contrast to CDMS and Edelweiss which do its active 
	background rejection by looking for ionisation and phonons, 
	CRESST II plans to use simultaneous detection of light and phonons.
	CDMS II plans to use 42 detectors with a total mass of 6.8\,kg 
	of Ge by 2006 in the Soudan mine, CRESST II plans 
	to have $\sim$10\,kg of CdWO$_4$ in the Gran Sasso in some future. 
	The cryogenic experiments are, however, 
	operating at present only 600\,g of detectors 
	{\it or less}, after a decade of development. 
	CDMS has collected only 10.6\,kg\,d 
	of data over this time 
	(in 1999, since then {\it no} measurement)%
\cite{Gait-Nanp2000}, 
	Edelweiss only 4.53\,kg\,d%
\cite{Edel01}.
	Therefore, they 
	may have severe problems to expand their 
	small detector masses to several tens of kg or better 100\,kg, 
	as required for modulation search.
	This means that although e.g. CDMS II may reach a future 
	sensitivity in an exclusion plot as shown in Fig.%
\ref{fig:Bedn-Wp2000}b,
	it will not be able to look for the modulation signal. 
	A general problem in the present stage still seems 
	to be the reproducibility of the highly complicated cryo detectors. 
	In spite of this, phantasy is large enough, 
	to dream already about 1\,ton cryo detectors systems%
\cite{Gait-Nanp2000}.

	Other far future projects are the superheavy droplet detectors 
	PICASSO/SIMPLE%
\cite{DRIFT01}.
	They are working at present on a scale of 15 and 50\,g detectors. 
	Their idea is to use 10-100$\mu$m diameter 
	droplets of volatile C$_4$F$_{10}$, 
	C$_3$F$_8$, ... in metastable superheated condition 
	and to choose critical energy and radius such 
	that only nuclear recoils can trigger a phase transition, 
	but not $\gamma$ and $\beta$ particles. 
	The acoustic signal of the explosive bubble formation 
	will be observed. The expected sensitivity of a 1\,ton 
	module for spin-dependent WIMP-nucleon interaction 
	is shown in Figure 
\ref{fig:Bedn-Wp2000}b
	- as SDD 1\,ton (for an assumed U/Th contamination 
	of 10$^{-15}$\,g/g - U/Th $\alpha$-emitters can cause recoil events!).
	A drawback is that these detectors cannot 
	measure energy spectra of WIMPs.

	A very promising project which would yield 
	a nice signal identification, is DRIFT in the Boulby mine. 
	It is aiming at looking for the diurnal directional modulation (Fig.%
\ref{fig:HDMS-DRIFT}b).
	The idea is to detect tracks of nuclear 
	recoils in a TPC with Xe(Ar) by a multiwire read-out.
	A 1\,m$^3$ prototype (1.5\,kg Xe) is under construction. 
	It is seen as first component of the full 10\,m$^3$ DRIFT experiment%
\cite{DRIFT01,Martof00}.
	
		The ZEPLIN project uses scintillation and electro-luminescence 
	in two-phase xenon. Plans for ZEPLIN II 
	are 0.01 - 0.1 counts/kg\,d and 20\,kg of Xe. 
	While still waiting for results of ZEPLIN I, 
	plans are already discussed for a ZEPLIN IV%
\cite{Cline01}.	

	To return to the more 'earth-bound' projects: 
	DAMA will extend their mass to 250\,kg, 
	and plans to start operation in summer 2002%
\cite{DAMA01}.
	Also the NAIAD project (Boulby mine) plans 
	to use NaI - 40-100\,kg of NaI in a liquid scintillator 
	Compton veto%
\cite{Spooner-DM0Engl}.
	The projected NAIAD limits for 100\,kg\,y exposure 
	are 0.1\,c/\,kg\,d. 
	Because of the large mass it will be able to look for modulation.

\begin{figure}[h]
\includegraphics*[width=75mm, height=60mm]
{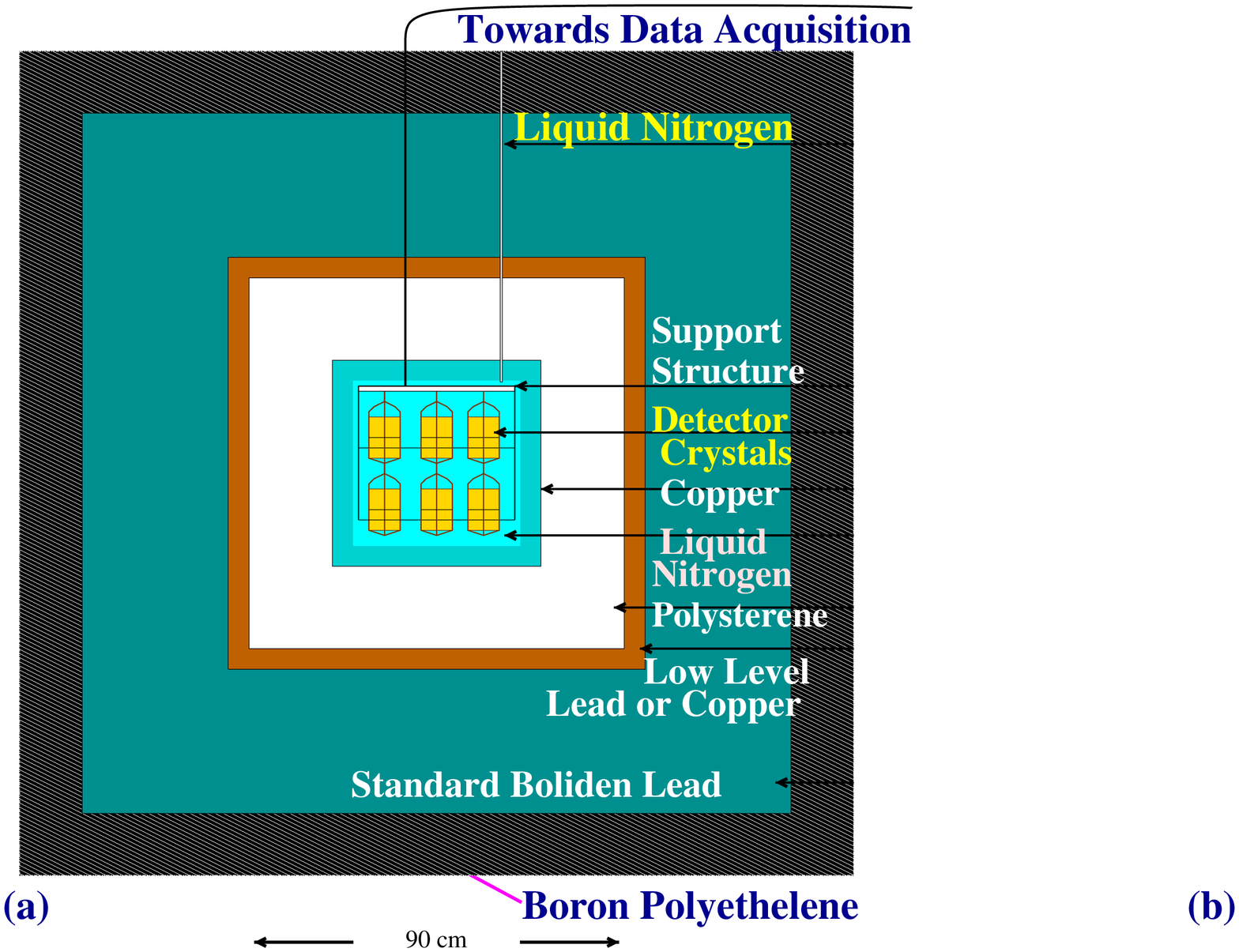}
\hspace{-0.3cm}
\includegraphics*[scale=0.25]{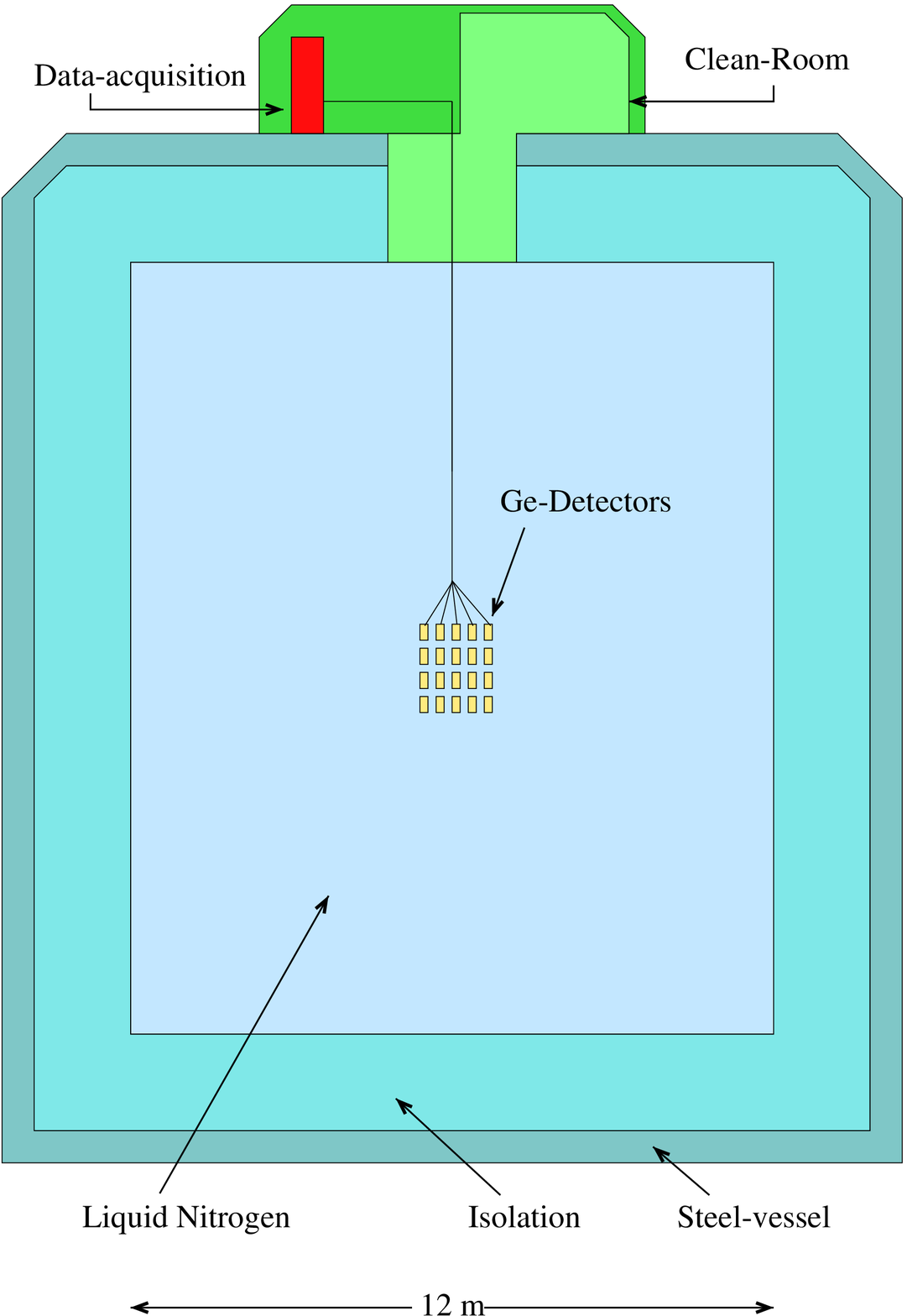}
\caption[]{
       (a): Conceptual design of the Genius TF. Up
	to 14 detectors will be housed in the inner detector chamber, 
	filled with liquid nitrogen. As a first shield 5 cm 
	of zone refined Germanium, or extremely low-level copper w
	ill be used.
	Behind the 20 cm of polystyrene isolation another 35 cm 
	of low level lead
	and a 15 cm borated polyethylene shield will complete the setup.
	(b): GENIUS - 100\,kg of Ge detectors 
	are suspended in a large liquid nitrogen tank (see%
\cite{KK-InJModPh98,GEN-prop,KK60Y}).
\label{fig:genius_tf_scheme}}
\end{figure}

\begin{figure}[h]

\vspace{-0.8cm}
\centering{
\includegraphics*[width=75mm, height=47mm]{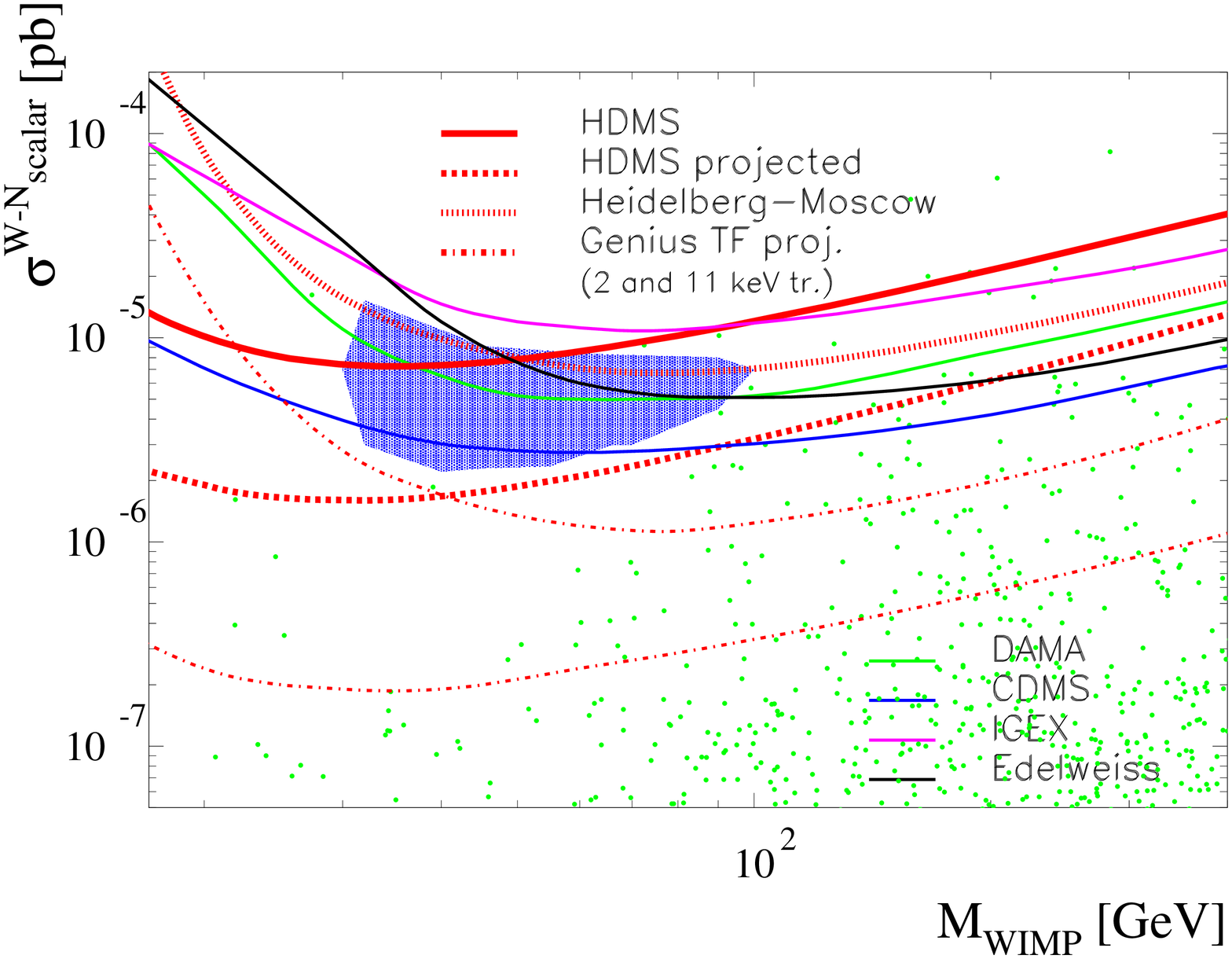}}

\vspace{-0.3cm}
\caption{The potential of GENIUS-TF for dark matter search. 
	  WIMP-nucleon cross section limits as a function of the WIMP
  	mass for spin-independent interactions. 
  	The solid lines are current limits 
	of the HEIDELBERG-MOSCOW experiment,  
  	the HDMS, 
	the DAMA  
	and the CDMS experiments.
  	The dashed curves are the expectation for HDMS,
	and for Genius-TF
  	with an energy threshold of 11\,keV and 2\,keV (no tritium
  	contamination) respectively, and a 
  	background index of 2\,events/kg\,y\,keV below 50\,keV.
  	The filled contour represents the 
	evidence region of the DAMA experiment
\label{fig:Upper-Gen-TF}}
\end{figure}

	The HDMS experiment and the GENIUS-TF experiment%
\cite{KK-GeTF-MPI,GenTF-0012022}
	aim at probing the DAMA evidence (see Fig.%
\ref{fig:Upper-Gen-TF}).
	GENIUS-TF consisting of 40\,kg of Ge detectors 
	in liquid nitrogen (Fig.%
\ref{fig:genius_tf_scheme}a)
	could also measure the modulation signal%
\cite{GenTF-0012022,Gen-01-Tom}. 
	Up to summer 2001, already 6 detectors of 2.5\,kg each, 
	with an extreme low-energy threshold of $\sim$500\,eV 
	have been produced. 
	A similar potential is aimed at by the GEDEON project%
\cite{MORALES},
	which plans to use 28 Ge diodes in one single cryostat.

	GENIUS-TF is already under installation in the Gran-Sasso 
	laboratory and should start operation by end of 2002.%

	The probably most far reaching project is GENIUS%
\cite{KK60Y,GEN-prop}.
	Since it is based on conventional techniques, 
	using Ge detectors in liquid nitrogen, 
	is may be realized in the most straightforward way.

	GENIUS would already in a first step, with 100\,kg of 
		{\it natural} Ge detectors in three years of measurement, 
	cover a significant part of the 
		SUSY parameter space for prediction of neutralinos 
		as cold dark matter 
(Figs.~\ref{fig:Bedn-Wp2000},\ref{fig:Ell-New}). 
	For this purpose the background in the energy range 
		$< 100$\,keV has to be reduced to 
		$10^{-2}$ (events/\,kg\,y\,keV). 
	At this level solar neutrinos as source of background 
	are still negligible. 
	Of particular importance is to shield the detectors 
	during production (and transport) to keep the background 
	from spallation by cosmic rays sufficiently  
	low (for details see%
\cite{KK-NOW00,KK-NOON00,KK-LowNu2}).

     The sensitivity of GENIUS for Dark Matter corresponds to 
	     that obtainable with a 1\,km$^3$ AMANDA detector for 
	     {\it indirect} detection (neutrinos from annihilation 
	     of neutralinos captured at the Sun) (see%
\cite{Eds99}). 
	Interestingly both experiments would probe different neutralino 
	compositions: GENIUS mainly gaugino-dominated neutralinos, 
	AMANDA mainly neutralinos with comparable gaugino and 
	Higgsino components (see Fig.\,38 in%
\cite{Eds99}).

\vspace{-0.3cm}
\subsection{Hot Dark Matter Search}

	According to the recent indication for the neutrinoless 
	mode of double beta decay%
\cite{KK01-Ev}, 
	(see section 2), 
	neutrinos should still play an important role 
	as hot dark matter in the Universe. 

	The effective mass has been determined to be%
\cite{KK01-Ev}
	$\langle m \rangle $= (0.05 - 0.84)\,eV at a 95$\%$ c.l.
	(best value 0.39\,eV) {\it including} 
	an uncertainty of $\pm$50$\%$ of the nuclear matrix elements.

\begin{figure}[h]
\centering{
\includegraphics*[height=6.2cm,width=12cm]
{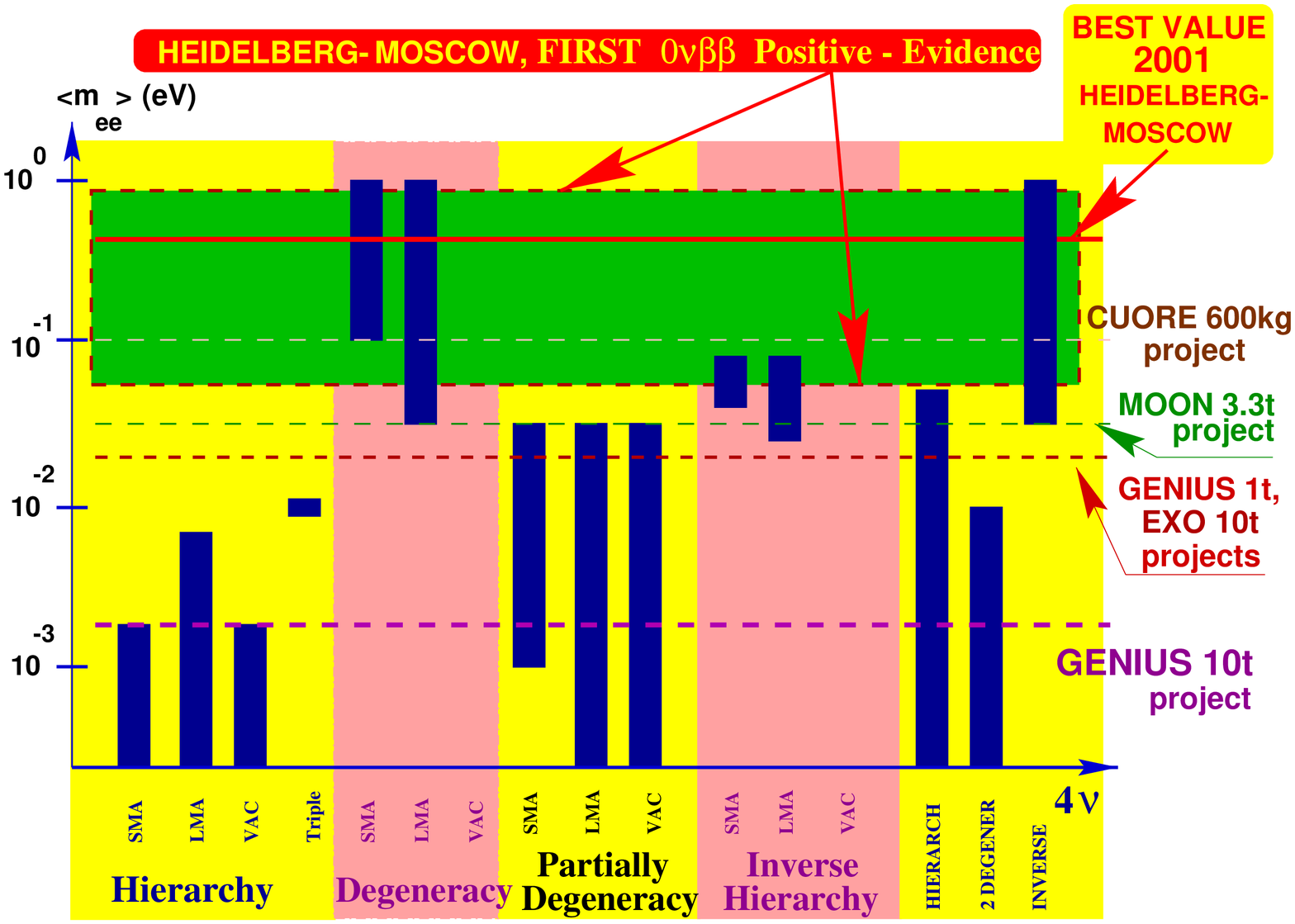}}
\caption[]{
	The impact of the evidence obtained for neutrinoless 
	double beta decay%
\cite{KK01-Ev}
	(best value of the effective neutrino mass 
	$\langle m \rangle$ = 0.39\,eV, 95$\%$ 
	confidence range (0.05 - 0.84)\,eV - 
	allowing already for an uncertainty of the nuclear 
	matrix element of a factor of $\pm$ 50$\%$,  
	on possible neutrino mass schemes. 
	The bars denote allowed ranges of $\langle m \rangle$ 
	in different neutrino mass scenarios, 
	still allowed by neutrino oscillation experiments.
	Hierarchical models are excluded by the 
	new \znbb~ decay result. Also shown are 
	the expected sensitivities 
	for the future potential double beta experiments 
	CUORE, MOON, EXO,   
	and the 1 ton and 10 ton project of GENIUS%
\cite{GEN-prop,KK-NOW00,KK-LowNu2,KK-NOON00}
	(from%
\cite{KK-Sark01-Ev}).}
 \label{fig:Jahr00-Sum-difSchemNeutr}
\end{figure}

	With the limit deduced for the effective neutrino mass,  
	the HEIDELBERG-MOSCOW experiment excludes several 
	of the neutrino mass 
	scenarios 
	allowed from present neutrino oscillation experiments
	(see Fig.%
\ref{fig:Jahr00-Sum-difSchemNeutr}) 
	- allowing mainly only for degenerate 
	and partially degenerate mass 
	scenarios and an inverse hierarchy 3$\nu$ - scenario
	(the latter being, however, strongly disfavored 
	by a recent analysis
	of SN1987A).   
	In particular hierarchical mass schemes 
	are excluded.

\begin{figure}[h]
\centering{
\includegraphics*[scale=0.35]
{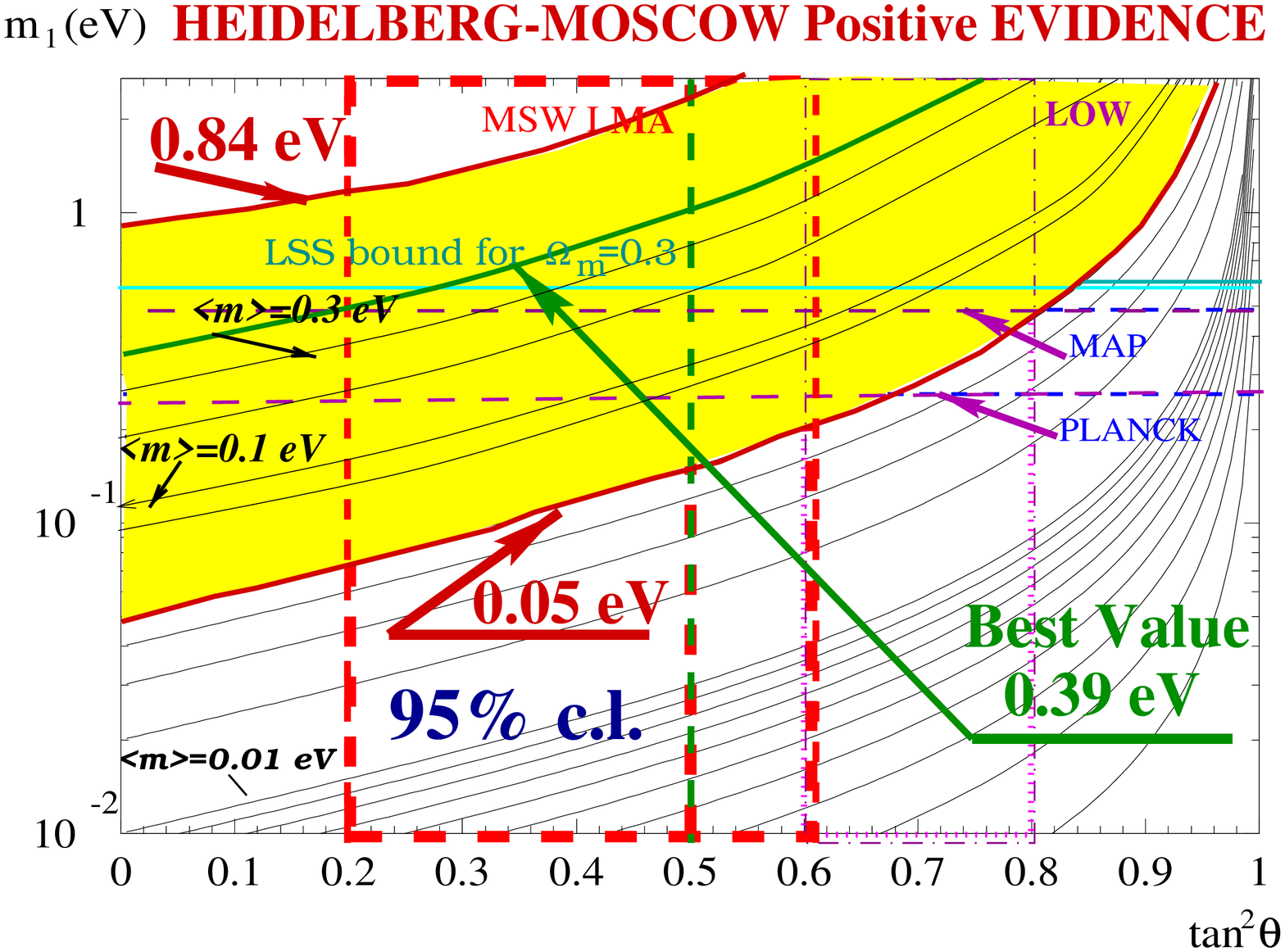}}

\vspace{-0.3cm}
\caption[]{
       Double beta decay observable 
	$\langle m\rangle$
	 and oscillation parameters: 
	 The case for degenerate neutrinos. 
	 Plotted on the axes are the overall scale of neutrino masses 
	 $m_0$ and mixing $\tan^2\, \theta^{}_{12}$. 
	 Also shown is a cosmological bound deduced from a fit of 
	 CMB and large scale structure%
\cite{Lop} 
	and the expected sensitivity of the satellite experiments 
	 MAP and PLANCK. 
	 The present limit from tritium $\beta$ decay of 2.2\,eV%
\cite{Trit00} 
	would lie near the top of the figure. 
     The range of 
	$\langle m\rangle$ 
	 fixed by the HEIDELBERG-MOSCOW experiment 
	is, in the case of small solar neutrino mixing, already in the 
	 range to be explored by MAP and PLANCK. 
\label{fig:Map-Planck}}
\end{figure}

	Assuming the degenerate scenarios to be realized in nature 
	we fix - according to the formulae derived in%
\cite{KKPS} - 
	the common mass eigenvalue of the degenerate neutrinos 
	to m = (0.05 - 3.4)\,eV. 
	Part of the upper range is already excluded by 
	tritium experiments, 
	which give a limit of m $<$ 2.2\,eV (95$\%$ c.l.)%
\cite{Trit00}.
	The full range can only  partly 
	(down to $\sim$ 0.5\,eV) be checked by future  
	tritium decay experiments,  
	but could be checked by some future $\beta\beta$ 
	experiments (see, e.g.%
\cite{KK60Y,KK-LowNu2,KK-NOON00}).
	The deduced best value for the mass 
	is consistent with expectations from experimental 
	$\mu ~\to~ e\gamma$
	branching limits in models assuming the generating 
	mechanism for the neutrino mass to be also responsible 
	for the recent indication for as anomalous magnetic moment 
	of the muon%
\cite{MaRaid01}.
	It lies in a range of interest also for Z-burst models recently 
	discussed as explanation for super-high energy cosmic ray events 
	beyond the GKZ-cutoff%
\cite{PW01-Wail99}.
	The sensitivity of the present result is already 
	in the range to be probed by the satellite 
	experiments MAP and PLANCK (Fig.%
\ref{fig:Map-Planck}).

\noindent
{\bf 4~~~~~Conclusion}

\vspace{0.2cm}
\noindent
	Dark matter search is presently one of the most exciting 
	fields of particle physics and cosmology. 
	Underground experiments at present only 
	marginally touch in their sensitivity the range 
	of present SUSY predictions for cold dark matter. 
	Of future experiments the GENIUS project has the best 
	prospects to cover a large part of the predicted range.  
	GENIUS will provide information complementary to future 
	collider search.
	This information is indispensable, even if LHC would 
	find supersymmetry, since in any case it still has to be shown 
	that SUSY particles indeed form the cold dark matter 
	in the Universe. 
	GENIUS will simultaneously be the most straightforward way 
	to fix the neutrino mass and the contributions 
	of neutrinos to hot dark matter with higher accuracy.

\small{
        
}

\end{document}